\documentclass[12pt]{article} 
\pdfoutput=1
\usepackage[authoryear]{natbib}
\usepackage{graphicx,epstopdf}
\usepackage{setspace}
\usepackage[margin=2.5cm]{geometry}
\usepackage{array}
\usepackage{hyperref}
\usepackage{xcolor}
\usepackage{amsfonts}
\usepackage{amsmath}
\usepackage{graphicx,lscape}
\usepackage{epstopdf}
\usepackage{rotating}
\usepackage{latexsym}
\usepackage{amssymb}

\usepackage[authoryear]{natbib}

\usepackage{pdflscape}
\usepackage{lscape}
\usepackage{graphicx}
\usepackage[utf8]{inputenc}
\usepackage[english]{babel}
\usepackage{enumitem}
\usepackage{subfig}
\usepackage[section]{placeins}
\usepackage{framed}

\usepackage{amsmath, amsthm, amssymb, amsfonts}

\newtheorem{corollary}{Corollary}
\newtheorem{proposition}{Proposition}
\newtheorem{lemma}{Lemma}
\newtheorem{remark}{Remark}

\newtheorem{assumption}{Assumption}

\usepackage{float}
\floatstyle{ruled}
\newfloat{algorithm}{tbp}{loa}
\providecommand{\algorithmname}{Algorithm}
\floatname{algorithm}{\protect\algorithmname}

\usepackage{ifpdf}
\ifpdf
\DeclareGraphicsExtensions{.pdf,.png,.jpg}
\else
\DeclareGraphicsExtensions{.eps}
\fi

\usepackage{natbib}
\setcitestyle{authoryear,open={(},close={)}}

\begin{document}
	
	\title{Identification in a Binary Choice Panel Data Model\\ with a Predetermined Covariate}
	\author{Stéphane Bonhomme, Kevin Dano, and Bryan S. Graham\thanks{\underline{Bonhomme}: Department of Economics, University of Chicago, 1126 E. 59th Street, Chicago, IL 60637, e-mail: [sbonhomme@uchicago.edu], web: \texttt{https://sites.google.com/site/stephanebonhommeresearch/}. \newline \underline{Dano}: Department of Economics, University of California - Berkeley, 530 Evans Hall \#3380, Berkeley, CA 94720-3880, e-mail: [kdano@berkeley.edu], web: \texttt{https://kevindano.github.io}. \newline  \underline{Graham}: Department of Economics, University of California - Berkeley, 530 Evans Hall \#3380, Berkeley, CA 94720-3880 and National Bureau of Economic Research, e-mail: [bgraham@econ.berkeley.edu], web: \texttt{http://bryangraham.github.io/econometrics/}. \newline We thank the Editor Laura Hospido and two anonymous reviewers for helpful comments. We are grateful to Isaiah Andrews, Manuel Arellano, Jes\'{u}s Carro, Bo Honor\'e, Martin Weidner, and the audiences at the conference in honor of Manuel Arellano at the Bank of Spain (July 2022) and other seminars for valuable comments that have improved the paper. All the usual disclaimers apply. Portions of the research reported here were undertaken while Bonhomme and Graham were visiting CEMFI with support from the Spanish State Research Agency under the María de Maeztu Unit of Excellence Programme (Project No: CEX2020-001104-M). The data (replication codes) will be available online. The authors have no conflict of interest to report.}} 
	
	\date{\today}
	
	\maketitle

	\begin{abstract}
		We study identification in a binary choice panel data model with a single \emph{predetermined} binary covariate (i.e., a covariate sequentially exogenous conditional on lagged outcomes and covariates). The choice model is indexed by a scalar parameter $\theta$, whereas the distribution of unit-specific heterogeneity, as well as the feedback process that maps lagged outcomes into future covariate realizations, are left unrestricted. We provide a simple condition under which $\theta$ is never point-identified, no matter the number of time periods available. This condition is satisfied in most models, including the logit one. We also characterize the identified set of $\theta$ and show how to compute it using linear programming techniques. While $\theta$ is not generally point-identified, its identified set is informative in the examples we analyze numerically, suggesting that meaningful learning about $\theta$ may be possible even in short panels with feedback. As a complement, we report calculations of identified sets for an average partial effect, and find informative sets in this case as well.       
	\end{abstract}
	
	
	
	\newpage
	
	\pagenumbering{arabic}
	\onehalfspacing
	
	\vspace{5mm}
	
	\section{Introduction}
	
	Empirical researchers utilizing panel data generally maintain the assumption that covariates are strictly exogenous: realized values of past, current, and future explanatory variables are independent of the time-varying structural disturbances or ``shocks''.\footnote{Dependence between the covariates and the time-invariant heterogeneity -- the so-called ``fixed effects'' -- is, of course, allowed.} In many settings this assumption is unrealistic. If the covariate is a policy, choice or dynamic state variable, then agents may adjust its level in response to past shocks (as when, for example, a firm adjusts its current capital expenditures in response to past productivity shocks).
	
	When strict exogeneity is untenable, \emph{sequential exogeneity} -- sometimes called \emph{predeterminedness} -- may be palatable. A predetermined covariate varies independently of current and future time-varying shocks, but general \emph{feedback}, or dependence on past shocks, is allowed. Assumptions of this type play an important role in, for example, production function estimation (\citealp{Olley_Pakes_EM96}, \citealp{Blundell_Bond_ER00}).
	
	In two seminal papers, \cite{arellano1991some} and \cite{Arellano_Bover_JE95}, Manuel Arellano and his collaborators presented foundational analyses of questions of identification, estimation, efficiency and specification testing in linear panel data models with feedback. Today such models are both well-understood and widely-used (see \cite{arellano2003panel} for a textbook review). 
	
	In contrast, the properties of nonlinear models with feedback are much less well-understood. In this paper we study binary choice. Most existing work in this area focuses on the case where the covariate is either strictly exogenous or a lagged outcome. Under strict exogeneity, \citet{rasch1960studies} and \citet{andersen1970asymptotic} show that the coefficient on the covariate is point-identified using two periods of data when shocks are logistic.
	\citet{chamberlain2010binary} provides conditions under which the logit case is the only one admitting point-identification with two periods (\citet{davezies2020fixed} provide extensions of this result to the case of $T>2$). In the dynamic case, where the covariate is a lagged outcome, \cite{Cox_JRSS58}, \citet{chamberlain1985heterogeneity} and \citet{honore2000panel} derive conditions for point-identification of the coefficient on the lagged outcome in the logit case, while \citet{honore2006bounds} show how to compute bounds on coefficients for probit and other models.

	Results for binary choice panel models with predetermined covariates are limited. \cite{Chamberlain_JOE2022} studies identification and semiparametric efficiency bounds in a class of non-linear panel data models with feedback; he provides both positive and negative results. In an hitherto unpublished section of an early draft of that paper (\citealp{Chamberlain_1993}), he proves that the coefficient on a lagged outcome is not point-identified in a dynamic logit model when only three periods of outcome data are available. \citet{Arellano_Carrasco_JOE2003} and \citet{honore2002semiparametric} study binary choice models with predetermined covariates. \citet{Arellano_Carrasco_JOE2003} assume that the dependence between the time-invariant heterogeneity and the covariates is fully characterized by its conditional mean given current and lagged covariates. \citet{honore2002semiparametric} assume that one of the covariates is independent of the individual effects conditional on the other covariates. In a recent contribution, \citet{pigini2022conditional} show that one can accommodate specific forms of feedback while maintaining point-identification in binary choice models with pretermined covariates.\footnote{In this paper we focus on panel data with a fixed number $T$ of time periods. The large-T literature has also considered models with dynamics and feedback, see for example \citet{carro2007estimating}, \citet{hahn2002asymptotically}, and \citet{fernandez2009fixed}.}   
	
	In what follows we pose two questions. First, under what conditions is the coefficient on a predetermined covariate in a binary choice panel data model point-identified? Second, when the coefficient is only set-identified, how extreme is the failure of point-identification; i.e., what is the width of the identified set? 
	
	Our analyses leave the dependence between the (time-invariant) unit-specific heterogeneity and the covariates unrestricted. We focus on the special case of a single binary predetermined covariate, leaving the feedback process from lagged outcomes, covariates and the unit-specific heterogeneity onto future covariate realizations fully unrestricted. This is a substantial relaxation of the strict exogeneity assumption.
	
	Regarding point-identification, we provide a simple condition on the model which guarantees that point-identification fails when $T$ periods of data are available (and $T$ is fixed). The condition is satisfied in most familiar models of binary choice, including the logit one. This finding contrasts with the prior work on logit models cited above, where point-identification typically holds for a sufficiently long panel. As a notable exception, the exponential binary choice model introduced by \citet{Al-Sadoon_et_al_ER2017} does not satisfy our condition. In fact, point-identification holds in that case.
	
	Regarding identified sets, we first show that sharp bounds on the coefficient can be computed using linear programming techniques. Our method builds on \citet{honore2006bounds}, however, in contrast to their work, we allow for heterogeneous feedback. While the regressor coefficient is our main target parameter, we also derive the identified set for an average partial effect. This set can be computed using linear programming techniques as well.
	
	Second, we numerically compute examples of identified sets. We find that, relative to the strictly exogenous case, allowing for a predetermined covariate tends to increase the width of the identified set. However, our calculations also suggest that the identified set can remain informative under predeterminedness, even in panels with as few as two periods, for both the coefficient and the average partial effect. Finally, as is true under strict exogeneity, the widths of the identified sets decrease quickly as the number of periods increases. These observations are based upon sets computed under a particular data generating processe (DGP). It is possible that identified sets may be larger under certain types of feedback.

	The outline of the paper is as follows. In Section \ref{sec_model} we present the model. In Section \ref{sec_id_T2} we provide a condition that implies that the common parameter in this model is not point-identified when $T=2$. In Section \ref{sec_id_T} we show that our condition implies failure of point-identification for all (finite) $T$. In Section \ref{sec_idset} we show how to compute identified sets on coefficients and average partial effects, and we report the results of a small set of numerical illustrations. In Section \ref{sec_restrict} we describe potential restrictions one could impose on the feedback process. These restrictions may restore point-identification or shrink the identified set.
	We conclude in Section \ref{sec_conc}. Proofs are contained in the appendix. Lastly, \href{https://github.com/kevindano/Bonhomme-Dano-Graham-SERIES}{replication codes} are available as supplementary material.
	
	\section{The model\label{sec_model}}
	
	Available to the econometrician is a random sample of $n$ units, each of which is followed for $T\geq2$ time periods. We focus on short panels, and keep $T$ fixed. The sampling process asymptotically reveals the joint distribution of $(X_1,\ldots,X_T,Y_1,\ldots,Y_T)$.
	
	For any sequence of random variables $Z_{t}$ and any non-stochastic sequence $z_t$, we use the shorthand notation
	$Z^{t:t+s}=(Z_{t}',...,Z_{t+s}')'$ and $z^{t:t+s}=(z_{t}',...,z_{t+s}')'$. In addition, we simply denote $Z^t=Z^{1:t}$ and $z^{t}=z^{1:t}$ when the subsequence starts in the first period.
	
	Let $Y_{it}\in\{0,1\}$ and $X_{it}\in\{0,1\}$ denote a binary outcome and a binary covariate, respectively. We assume that 
	$$\Pr(Y_{it}=1\,|\, Y_{i}^{t-1},X_{i}^{t},\alpha_i;\theta)=F(\theta X_{it}+\alpha_i), \quad t=1,\ldots,T,$$
	where $\alpha_i\in{\cal{S}}\subset \mathbb{R}$ is a scalar individual effect, $F(\cdot)$ is a known differentiable cumulative distribution function, and $\theta\in\Theta$ is a scalar parameter. 
	
	Let $\pi_{x_1}(\alpha)$
	denote the \emph{distribution of heterogeneity} given the initial condition $X_1=x_1$; i.e., the distribution of $\alpha_i\,|\, X_{i1}$. We leave this distribution unrestricted on ${\cal{S}}$. When ${\cal{S}}$ is a discrete subset of the real line, $\pi_{x_1}(\alpha)$ belongs to the unit simplex on ${\cal{S}}$, however it is otherwise unrestricted.  We denote as $\Pi$ the collection of all $\pi_{x_1}(\alpha)$, for all $x_1\in\{0,1\}$ and $\alpha\in{\cal{S}}$.

	For each $t\geq 2$, let
	$$\Pr\left(X_{it}=1\,|\, Y_{i}^{t-1}=y^{t-1},X_{i}^{t-1}=x^{t-1},\alpha_i=\alpha\right)=G^t_{y^{t-1},x^{t-1}}(\alpha), \quad t=2,\ldots,T,$$
	denote the \emph{feedback process} through which lagged outcomes, past covariates and heterogeneity affect the current covariate. We leave this distribution unrestricted as well. We denote as $G\in {\cal{G}}_T$ the collection of all $G^t_{y^{t-1},x^{t-1}}(\alpha)$, for all $t\in\{2,...,T\}$, $y^{t-1}\in\{0,1\}^{t-1}$, $x^{t-1}\in\{0,1\}^{t-1}$, and $\alpha\in{\cal{S}}$.

	The (integrated) likelihood function conditional on the first period's covariate is
	\begin{align}\Pr\left(Y_{i}^T=y^T,X_{i}^{2:T}=x^{2:T}\,|\, X_{i1}=x_1\right)&=\int_{{\cal{S}}}\, \underset{\text{outcomes}}{\underbrace{\prod_{t=1}^T F(\theta x_t+\alpha)^{y_t}[1-F(\theta x_t+\alpha)]^{1-y_t}}}\notag\\
		&\quad\quad\quad\times \underset{\text{feedback}}{\underbrace{\prod_{t=2}^T G^t_{y^{t-1},x^{t-1}}(\alpha)^{x_t}[1-G^t_{y^{t-1},x^{t-1}}(\alpha)]^{1-x_t}}}\notag\\
		& \quad\quad\quad\quad\quad\quad\times \underset{\text{heterogeneity}}{\underbrace{\pi_{x_1}(\alpha)}}d\mu(\alpha),\label{lik_eq}\end{align}
	for some (discrete or continuous) measure $\mu$ on ${\cal{S}}$. 
	
	A key feature of a model with predetermined covariates is the dependence of the feedback process on lagged outcomes, as reflected in the dependence of $G^t$ on $y^{t-1}$ in (\ref{lik_eq}). When this dependence is ruled out, the covariate is strictly exogenous, and the likelihood function simplifies.\footnote{Under strict exogeneity, the likelihood function factors as
		\begin{align*}\Pr\left(Y_{i}^T=y^T,X_{i}^{2:T}=x^{2:T}\,|\, X_{i1}=x_1\right)&=\bigg[\int_{{\cal{S}}}\, \prod_{t=1}^T F(\theta x_t+\alpha)^{y_t}[1-F(\theta x_t+\alpha)]^{1-y_t}\pi_{x^T}(\alpha)d\mu(\alpha)\bigg]\\
			&\quad \quad \quad \times \Pr\left(X_{i}^{2:T}=x^{2:T}\,|\, X_{i1}=x_1\right),\end{align*}
		where $\pi_{x^T}(\alpha)$ denotes the distribution of heterogeneity given all periods' covariates $x_1,...,x_T$.
	} Dynamic responses of covariates to lagged outcome realizations are central to many economic models, including those where $X_{it}$ is a choice variable, policy, or a dynamic state variable.

	For any $(\theta,\pi,G)\in\Theta\times\Pi\times{\cal{G}}_T$, and any $(y^T,x^{2:T})\in \{0,1\}^{2T-1}$, let $Q_{x_1}(y^T,x^{2:T};\theta,\pi,G)$ denote the right-hand side of (\ref{lik_eq}). Moreover, let $Q_{x_1}(\theta,\pi,G)$ denote the $2^{2T-1}\times 1$ vector collecting all those elements, for all $(y^T,x^{2:T})\in \{0,1\}^{2T-1}$. Finally, let $Q(\theta,\pi,G)$ denote the  $2^{2T}\times 1$ vector stacking $Q_{1}(\theta,\pi,G)$ and $Q_{0}(\theta,\pi,G)$. For a given (population) $(\theta,\pi,G)\in \Theta\times\Pi\times{\cal{G}}_T$, we define the \emph{identified set} of $\theta$ as
	\begin{equation}\Theta^I=\left\{\widetilde\theta\in\Theta\,:\, \exists (\widetilde\pi,\widetilde G)\in \Pi\times{\cal{G}}_T\,:\,Q(\widetilde\theta,\widetilde\pi,\widetilde G)=Q(\theta,\pi,G)\right\}.\label{eq_Theta_I}\end{equation}
	The set in \eqref{eq_Theta_I} includes all $\widetilde\theta\in\Theta$ where, for that $\widetilde\theta$, it is possible to find a heterogeneity distribution $\widetilde\pi\in\Pi$, \emph{and} a feedback process $\widetilde G\in \cal{G}_T$, such that the resulting conditional likelihood assigns the same probability to each of the $2^{2T-1}$ possible data outcomes as the true one (given both $X_{i1}=0$ and $X_{i1}=1$).
	
	In the first part of the paper, we provide conditions on the model under which $\Theta^I$ is not a singleton. This corresponds to cases where $\theta$ is not point-identified. In the second part of the paper, we report numerical calculations of $\Theta^I$ under particular DGPs.
	
	Our focus on $\theta$ is motivated by the extensive literature on the identification of coefficients in binary choice models. However, in applications, average effects may also be of interest. In the second part of the paper, we will also report numerical calculations of identified sets for an average partial effect associated with a change in the binary predetermined covariate.

	\section{Failure of point-identification in two-period panels \label{sec_id_T2}}
	
	We first present an analysis of point-identification in the two-period case, since this leads to simple and transparent calculations. In the next section, we will then generalize this result to accommodate $T\geq 2$ periods. 
	
	\subsection{Assumptions and result}

	To keep the formal analysis simple, in this section and the next we assume that $\alpha_i$ takes a finite number of values, with known support points.
	
	\begin{assumption}\label{ass_disc}
		${\cal{S}}=\{\underline{\alpha}_1,...,\underline{\alpha}_K\}$, where $\underline{\alpha}_1,...,\underline{\alpha}_K$ are known, and $\mu= \sum_{k=1}^K\delta_{\underline{\alpha}_k}$, where $\delta_{\alpha}$ denotes the Dirac measure at $\alpha$.
	\end{assumption}
	
	Assumption \ref{ass_disc} makes the model fully parametric. However this is not a limitation as our aim in this section and the next is to derive conditions under which point-identification \emph{fails}. The conditions we provide will require sufficiently many support points.\footnote{The analysis is essentially unchanged if one instead assumes that $\mu=\sum_{k=1}^K\lambda_k\delta_{\underline{\alpha}_k}$, for some $\lambda_k>0$.}
	
	We rely on the parameterization given by the $2(K-1)\times 1$ vector $\pi=(\pi_1',\pi_0')'$, where, for all $x_1\in\{0,1\}$,
	$\pi_{x_1}=(\pi_{x_1}(\underline{\alpha}_1),\ldots,\pi_{x_1}(\underline{\alpha}_{K-1}))'$ and $\pi_{x_1}(\underline{\alpha}_K)=1-\sum \limits_{k=1}^{K-1}\pi_{x_1}(\underline{\alpha}_k)$. The vector $\pi\in\Pi$ is unrestricted, except for the fact that $\pi_{x_1}(\alpha)$, for $\alpha\in{\cal{S}}$, belongs to the unit simplex. This parameterization handles the fact that probability mass functions sum to one. 
	
	We next impose the following assumption on the population parameters. 
	
	\begin{assumption}\label{ass_1}
		$\theta\in\Theta$, $\pi\in\Pi$, and $G\in{\cal{G}}_T$ are all interior, and $F(\theta x+\alpha)\in(0,1)$ for all $x\in\{0,1\}$ and $\alpha\in {\cal{S}}$.
	\end{assumption}
	
	Assumption \ref{ass_1} places restrictions on the underlying parametric binary choice model and heterogeneity distribution. It rules out heterogeneity distributions that induce a point mass of ``stayers'' (i.e., units with such extreme values of $\alpha$ that they either always take the binary action or they never do).\footnote{In some microeconometric datasets a substantial fraction of units never alter their value of $X_t$. For example, in \cite{CardEM1996} few workers join or leave a union during the sample period.} Assumption \ref{ass_1} also rules out the ``staggered adoption'' design common in difference-in-differences analyses. Exploring the implications of non-interior feedback processes is left for future work.
	
	Finally, we assume that the parameter point is regular in the sense of \citet{rothenberg1971identification}.

	\begin{assumption}\label{ass_2}
		$(\theta,\pi,G)$ is a regular point of the Jacobian matrix $\nabla Q(\theta,\pi,G)$, in the sense that the rank of $\nabla Q(\widetilde{\theta},\widetilde{\pi},\widetilde{G})$ is constant for all $(\widetilde{\theta},\widetilde{\pi},\widetilde{G})$ in an open neighborhood of $(\theta,\pi,G)$.
	\end{assumption}
	
	The assumption of regularity is standard in the literature on the identification of parametric models (\citealp{rothenberg1971identification}). If $F(\cdot)$ is analytic, the irregular points of $\nabla Q(\theta,\pi,G)$ (i.e., the points $(\theta,\pi,G)$ such that Assumption \ref{ass_2} is not satisfied) form a set of measure zero (\citealp{bekker2001identification}). Thus, Assumption \ref{ass_2} is satisfied almost everywhere in the parameter space in many binary choice models, including the probit and logit ones. 
	
	We aim to provide a simple condition under which point-identification of $\theta$ fails when $T=2$. We start by observing that, when $T=2$, the $2^{2T-1}=8$ model outcome probabilities given $X_{i1}=x_1$ are
	\begin{align*}
		Q_{x_1}(\theta,\pi,G)&=\left(\begin{array}{c}\Pr(Y_{i2}=1,X_{i2}=1,Y_{i1}=1\,|\, X_{i1}=x_1;\theta,\pi,G)\\
			\Pr(Y_{i2}=1,X_{i2}=1,Y_{i1}=0\,|\, X_{i1}=x_1;\theta,\pi,G)\\
			\Pr(Y_{i2}=1,X_{i2}=0,Y_{i1}=1\,|\, X_{i1}=x_1;\theta,\pi,G)\\
			\Pr(Y_{i2}=1,X_{i2}=0,Y_{i1}=0\,|\, X_{i1}=x_1;\theta,\pi,G)\\
			\Pr(Y_{i2}=0,X_{i2}=1,Y_{i1}=1\,|\, X_{i1}=x_1;\theta,\pi,G)\\
			\Pr(Y_{i2}=0,X_{i2}=1,Y_{i1}=0\,|\, X_{i1}=x_1;\theta,\pi,G)\\
			\Pr(Y_{i2}=0,X_{i2}=0,Y_{i1}=1\,|\, X_{i1}=x_1;\theta,\pi,G)\\
			\Pr(Y_{i2}=0,X_{i2}=0,Y_{i1}=0\,|\, X_{i1}=x_1;\theta,\pi,G)
		\end{array}\right),
	\end{align*}
	which, given the structure of the model, coincide with
	\begin{align}
		Q_{x_1}(\theta,\pi,G)
		&=\left(\begin{array}{c}\int_{{\cal{S}}}F(\theta +\alpha) G^2_{1,x_1}(\alpha)F(\theta x_1+\alpha)\pi_{x_1}(\alpha)d\mu(\alpha)\\
			\int_{{\cal{S}}} F(\theta +\alpha) G^2_{0,x_1}(\alpha)[1-F(\theta x_1+\alpha)]\pi_{x_1}(\alpha)d\mu(\alpha)\\
			\int_{{\cal{S}}}	F(\alpha) [1-G^2_{1,x_1}(\alpha)]F(\theta x_1+\alpha)\pi_{x_1}(\alpha)d\mu(\alpha)\\
			\int_{{\cal{S}}}	F(\alpha) [1-G^2_{0,x_1}(\alpha)][1-F(\theta x_1+\alpha)]\pi_{x_1}(\alpha)d\mu(\alpha)\\
			\int_{{\cal{S}}}	\left[1-F(\theta +\alpha)\right] G^2_{1,x_1}(\alpha)F(\theta x_1+\alpha)\pi_{x_1}(\alpha)d\mu(\alpha)\\
			\int_{{\cal{S}}}	\left[1-F(\theta +\alpha)\right] G^2_{0,x_1}(\alpha)[1-F(\theta x_1+\alpha)]\pi_{x_1}(\alpha)d\mu(\alpha)\\
			\int_{{\cal{S}}}	\left[1-F(\alpha)\right][1- G^2_{1,x_1}(\alpha)]F(\theta x_1+\alpha)\pi_{x_1}(\alpha)d\mu(\alpha)\\
			\int_{{\cal{S}}}	\left[1-F(\alpha)\right][1- G^2_{0,x_1}(\alpha)][1-F(\theta x_1+\alpha)]\pi_{x_1}(\alpha)d\mu(\alpha)
		\end{array}\right).\label{eq_Q_T2}
	\end{align}

	With this notation in hand we present the following lemma. 
	
	\begin{lemma}\label{lem_1}
		
		Let $T=2$. Suppose that Assumptions \ref{ass_disc}, \ref{ass_1} and \ref{ass_2} hold, and that $\theta$ is point-identified. Then, there exists $x_1\in \{0,1\}$ and a non-zero function $\phi_{x_1}:\{0,1\}^3\rightarrow \mathbb{R}$ such that: 
		
		(i) for all $\alpha\in {\cal{S}}$ and $y_1\in\{0,1\}$,
		\begin{align}&\sum_{y_2=0}^1\phi_{x_1}(y_1,y_2,1)F(\theta +\alpha)^{y_2}[1-F(\theta +\alpha)]^{1-y_2}=\sum_{y_2=0}^1\phi_{x_1}(y_1,y_2,0)F(\alpha)^{y_2}[1-F(\alpha)]^{1-y_2};\notag\\\label{eq_1}\end{align}
		
		(ii) for all $\alpha\in {\cal{S}}$ and $x_2\in\{0,1\}$,
		\begin{align}&\sum_{y_2=0}^1\sum_{y_1=0}^1\phi_{x_1}(y_1,y_2,x_2)F(\theta x_2+\alpha)^{y_2}[1-F(\theta x_2+\alpha)]^{1-y_2}F(\theta x_1+\alpha)^{y_1}[1-F(\theta x_1+\alpha)]^{1-y_1}=0.\notag\\\label{eq_2} 
		\end{align}

	\end{lemma}

	The proof of Lemma \ref{lem_1} exploits the fact that, if $\theta$ is point-identified, then it is also locally point-identified. Together with the assumption that the parameter is regular, this allows us to apply a result of \cite{bekker2001identification} regarding the identification of subvectors, which
	guarantees the existence of some $x_1\in\{0,1\}$ such that
	$\nabla_{\theta'} Q_{x_1}$ does not belong to the range of the matrix $\begin{bmatrix}
		\nabla_{\pi_{x_1}'} Q_{x_1} &\nabla_{G_{x_1}'} Q_{x_1}
	\end{bmatrix}$. We then show, using (\ref{eq_Q_T2}), that this implies the existence of $\phi_{x_1}\neq 0$ such that \eqref{eq_1} and \eqref{eq_2} hold. 
	
	When the population parameter $\theta$ is point-identified, Lemma \ref{lem_1} suggests a method-of-moments approach to estimation. In such settings, $\phi_{X_{i1}}\left(Y_{i1},Y_{i2},X_{i2}\right)$ will generally be a non-trivial function of $\theta$. Let $\phi_{X_{i1}}\left(Y_{i1},Y_{i2},X_{i2};\theta\right)$ be this function. Next, note that condition $\eqref{eq_1}$ in Lemma \ref{lem_1} corresponds to the conditional moment restriction \begin{equation}\mathbb{E}\left[\phi_{X_{i1}}\left(Y_{i1},Y_{i2},X_{i2};\theta\right)\,|\,X_{i1},X_{i2},Y_{i1},\alpha_{i}\right]=\mathbb{E}\left[\phi_{X_{i1}}\left(Y_{i1},Y_{i2},X_{i2};\theta\right)\,|\,X_{i1},Y_{i1},\alpha_{i}\right],\label{eq_feedback_rob}
	\end{equation} while  -- continuing to maintain $\eqref{eq_1}$ -- equation $\eqref{eq_2}$ implies the additional requirement that \begin{equation}\mathbb{E}\left[\phi_{X_{i1}}\left(Y_{i1},Y_{i2},X_{i2};\theta\right)\,|\,X_{i1},\alpha_{i}\right]=0.\label{eq_hetero_rob}\end{equation} 
Analog estimators in point-identified models with feedback, based on these observations, are explored in our companion paper \citep{BDG_WP}.
	
	This formulation clarifies that a necessary condition for point-identification of $\theta$ is the existence of a non-zero moment function, $\phi_{X_{i1}}\left(Y_{i1},Y_{i2},X_{i2};\theta\right)$, with a mean that is invariant to $X_{i2}$ given $\alpha_{i}$ and the past (i.e., the first period's covariate and outcome). Such a moment function is ``feedback robust'', in the sense that it remains valid across all possible feedback processes. This is the content of condition $\eqref{eq_1}$ in Lemma \ref{lem_1}, while $\eqref{eq_2}$ imposes a similar invariance to the distribution of unobserved heterogeneity.
	
	To show that point-identification fails, our focus here, we need to show that no such non-zero moment function exists. It turns out that there is a very simple condition for this in our model. Specifically, from Lemma \ref{lem_1} we obtain the following corollary.

	\begin{corollary}\label{coro_1}
		Let $T=2$. Suppose that Assumptions \ref{ass_disc}, \ref{ass_1} and \ref{ass_2} hold, and that $1$, $F(\alpha)$, and $F(\theta+\alpha)$, for $\alpha\in {\cal{S}}$, are linearly independent, then $\theta$ is not point-identified.\end{corollary}
	
	Corollary \ref{coro_1} shows that a necessary condition for identification of $\theta$ is that $1$, $F(\alpha)$, and $F(\theta+\alpha)$, for $\alpha\in {\cal{S}}$, are linearly dependent. This condition arises directly from condition (\ref{eq_1}), which requires the existence of a moment function that is robust to unknown feedback. Indeed, one can show that $1$, $F(\alpha)$, and $F(\theta+\alpha)$ are linearly dependent if and only if there exists a non-constant function $\phi$ such that 
	\begin{equation}\mathbb{E}\left[\phi(Y_{it},X_{it})\,|\, X_{it},\alpha_i\right]=\mathbb{E}\left[\phi(Y_{it},X_{it})\,|\, \alpha_i\right].\label{eq_feedback_robust}\end{equation}
	
	However, the condition that $1$, $F(\alpha)$, and $F(\theta+\alpha)$ be linearly dependent is restrictive, as we show in the next subsection.\footnote{While here we focus on a discrete ${\cal{S}}$ under Assumption \ref{ass_disc}, note that, when $\theta\neq 0$ and $F$ is strictly increasing on $\mathbb{R}$, $1$, $F(\alpha)$, and $F(\theta+\alpha)$, for $\alpha\in \mathbb{R}$, cannot be linearly dependent. If that were the case, then for some non-zero triplet $(A,B,C)$ we would have
		$AF(\theta+\alpha)+BF(\alpha)+C=0$ for all $ \alpha\in {\mathbb{R}}$.
		This would imply, by taking $\alpha\rightarrow \pm \infty$ that $C=0$ and $A+B=0$, which would then imply $A=B=C=0$ and contradict the assumption that $(A,B,C)$ is non-zero.}
	
	\begin{remark} \label{remark: sign_identification}
		Despite the negative result of Corollary \ref{coro_1}, the sign of $\theta$ is identified provided that Assumption \ref{ass_1} holds and $F(\cdot)$ is strictly increasing. Specifically, we show in Appendix \ref{appendix_section_sgntheta} that 
		$$		\mbox{ sign}(\theta)=\mbox{sign}\left(\mathbb{E}\left[Y_{i2}-Y_{i1}\,|\,X_{i1}=0\right]\right)=\mbox{sign}\left(\mathbb{E}\left[Y_{i1}-Y_{i2}\,|\,X_{i1}=1\right]\right).
		$$	
		
	\end{remark}
	
	\subsection{The logit model}
	
	Consider the logit model with a binary predetermined covariate, which corresponds to $F(u)=\frac{e^u}{1+e^u}$. In this case, the linear dependence condition of Corollary \ref{coro_1} requires that, for some non-zero triplet $(A,B,C)$,
	$$A\frac{e^{\theta +\alpha}}{1+e^{\theta +\alpha}}+B \frac{e^{\alpha}}{1+e^{\alpha}}+C=0,\quad \text{for all } \alpha\in{\cal{S}}.$$
	However, this implies
	$$Ae^{\theta }e^{\alpha}(1+e^{\alpha})+Be^{\alpha}(1+e^{\theta}e^\alpha)+C(1+e^{\alpha})(1+e^{\theta}e^\alpha)=0,\quad \text{for all } \alpha\in{\cal{S}},$$
	which is a quadratic polynomial equation in $e^{\alpha}$. Therefore, provided that there are $K\geq 3$ values in ${\cal{S}}$, this implies
	$$Ae^{\theta }+Be^{\theta }+Ce^{\theta }=0,\quad Ae^{\theta }+B+(1+e^{\theta})C=0,\quad C=0,$$
	which, provided that $\theta\neq 0$, entails
	$$A=B=C=0,$$
	contradicting the assumption that $(A,B,C)$ is non-zero.
	
	We have thus proved the following corollary.
	
	\begin{corollary}\label{coro_2}
		Consider the logit model with $T=2$. Suppose that Assumptions \ref{ass_disc}, \ref{ass_1} and \ref{ass_2} hold, that $\theta\neq 0$, and that ${\cal{S}}$ contains at least three points, then $\theta$ is not point-identified.
	\end{corollary}
	
	A precedent to Corollary \ref{coro_2} is given in the unpublished working paper by \citet{Chamberlain_1993} mentioned in the introduction. In the model he considers, $X_{it}=Y_{i,t-1}$ is a lagged outcome, and $T=2$ (hence, outcomes are observed for three periods). His model also includes an additional regressor: an indicator for period $t=2$.

	\subsection{The exponential model}
	
	Suppose now that, for $u\geq 0$, $F(u)=1-e^{-u}$. This corresponds to the exponential binary choice model of \citet{Al-Sadoon_et_al_ER2017}. Note that here the support of $F(\cdot)$ is a strict subset of the real line. In this case, letting 
	$$A=e^{\theta}, \,B=-1, \,C=1-e^{\theta},$$
	we have
	$$A[1-e^{-(\theta+\alpha)}]+B[1-e^{-\alpha}]+C=0.$$
	Hence the non point-identification condition of Corollary \ref{coro_1} is not satisfied in the exponential binary choice model.

	In fact, in this case (\ref{eq_1}) and (\ref{eq_2}) are satisfied for
	$$\phi_{x_1}(y_1,y_2,x_2;\theta)=(1-y_2)e^{\theta x_2}-(1-y_1)e^{\theta x_1},$$
	and $\theta$ satisfies the conditional moment restriction
	$$\mathbb{E}[\phi_{X_{i1}}(Y_{i1},Y_{i2},X_{i2};\theta)\,|\, X_{i1}]=0,$$
	that is,
	\begin{equation}\mathbb{E}[(1-Y_{i2})e^{\theta X_{i2}}-(1-Y_{i1})e^{\theta X_{i1}}\,|\, X_{i1}]=0.\label{eq_mom_expo}\end{equation} See \cite{wooldridge1997multiplicative} for several related results. Furthermore, one can show formally that $\theta$ is point-identified based on (\ref{eq_mom_expo}), see Appendix \ref{App_expo}.

	\section{Failure of point-identification in $T$-period panels for $T >2$\label{sec_id_T}}
	
	In this section we generalize our analysis to an arbitrary number of periods and state our main result.
	
	\subsection{Main result}

	The arguments laid out in the previous section extend to an arbitrary number of time periods, $T\geq2$. Indeed, using a similar strategy to the proof of Lemma \ref{lem_1} and proceeding by induction, we obtain the following lemma.
	
	\begin{lemma}\label{lem_2}
		
		Let $T\geq 2$. Suppose that Assumptions \ref{ass_disc}, \ref{ass_1} and \ref{ass_2} hold, and that $\theta$ is point-identified. Then, there exists $x_1\in \{0,1\}$ and a non-zero function $\phi_{x_1}:\{0,1\}^{2T-1}\rightarrow \mathbb{R}$ such that:
		
		(i) for all $\alpha\in {\cal{S}}$, $ s\in\{0,...,T-2\}$, $ y^{T-(s+1)}\in \{0,1\}^{T-(s+1)}$, $x^{T-(s+1)}\in \{0,1\}^{T-(s+1)}$,
		\begin{align}
			\sum_{y^{T-s:T}\in \{0,1\}^{s+1}} \phi_{x_1}(y^{T},x^{2:T})\prod_{t=T-s}^T F(\theta x_{t}+\alpha)^{y_t}[1-F(\theta x_{t}+\alpha)]^{1-y_t} \label{eq_1_general}
		\end{align}
		does not depend on $x^{T-s:T}$;
		
		(ii) for all $\alpha\in {\cal{S}}$ and $x^{2:T}\in\{0,1\}^{T-1}$,
		\begin{align}
			\sum_{y^{T}\in \{0,1\}^{T}} \phi_{x_1}(y^{T},x^{2:T})\prod_{t=1}^T F(\theta x_{t}+\alpha)^{y_t}[1-F(\theta x_{t}+\alpha)]^{1-y_t}=0.
			\label{eq_2_general}
		\end{align}
	\end{lemma}
	
	Similarly to Lemma \ref{lem_1}, Lemma \ref{lem_2} implies the existence of a moment function, with (generally) non-trivial dependence on $\theta$, which is ``feedback robust'', in the sense that, for all  $ s\in\{0,...,T-2\}$,
	$$ \mathbb{E}\left[\phi_{X_{i1}}(Y_i^{T},X_i^{2:T};\theta)\,|\, X_i^{T-s},Y_i^{T-(s+1)},\alpha_i\right]=\mathbb{E}\left[\phi_{X_{i1}}(Y_i^{T},X_i^{2:T};\theta)\,|\, X_i^{T-(s+1)},Y_i^{T-(s+1)},\alpha_i\right],$$
	while also requiring that
	$$ \mathbb{E}\left[\phi_{X_{i1}}(Y_i^{T},X_i^{2:T};\theta)\,|\, X_{i1},\alpha_i\right]=0.$$

	From Lemma \ref{lem_2} we obtain the following corollary, which we also prove by induction. This is our main result.
	
	\begin{corollary}\label{coro_3}
		Let $T\geq 2$. Suppose that Assumptions \ref{ass_disc}, \ref{ass_1} and \ref{ass_2} hold, and that $1$, $F(\alpha)$, and $F(\theta+\alpha)$, for $\alpha\in {\cal{S}}$, are linearly independent, then 
		$\theta$ is not point-identified.
	\end{corollary}

	\subsection{Logit model}
	
	Using that, when $\theta\neq 0$, $1$, $F(\alpha)$, and $F(\theta+\alpha)$, for $\alpha\in {\cal{S}}$, are linearly independent in the logit model, Corollary \ref{coro_3} implies that in the logit model with a binary predetermined covariate, $\theta$ is not point-identified irrespective of the number of time periods available.

	\begin{corollary}\label{coro_4}
		Consider the logit model with $T\geq 2$. Suppose that Assumptions \ref{ass_disc}, \ref{ass_1} and \ref{ass_2} hold, that $\theta\neq 0$, and that ${\cal{S}}$ contains at least three points, then $\theta$ is not point-identified. 
	\end{corollary}
	
	This non point-identification result contrasts with prior work on logit panel data models. Under strict exogeneity, \citet{rasch1960studies} and \citet{andersen1970asymptotic} have established that $\theta$ is point-identified under mild conditions on $X_{it}$ whenever $T\geq 2$. In the dynamic logit model when $X_{it}=Y_{i,t-1}$, \citet{Chamberlain_1993} shows that $\theta$ is not point-identified when $T=2$ (a result also obtained as an implication of Corollary \ref{coro_1}). However, \citet{chamberlain1985heterogeneity}, and \citet{honore2000panel} in a model with covariates, show that $\theta$ is point-identified under suitable conditions whenever $T\geq 3$.\footnote{Since in the dynamic logit model $X_{it}=Y_{i,t-1}$ is a lagged outcome, $T\geq 2$ (respectively, $T\geq 3$) requires that individual outcomes be available for at least three (resp., four) periods.} By contrast, Corollary \ref{coro_4} shows that, when the feedback process through which current covariates are influenced by lagged outcomes is unrestricted, the failure of point-identification is pervasive irrespective of $T$, despite the logit structure.

	\section{Characterizing identified sets\label{sec_idset}}
	
	The previous sections show that point-identification often fails in binary choice models with a predetermined covariate. In this section, we explore the degree of identification failure by presenting numerical calculations of the identified set $\Theta^I$ for specific parameter values. In the last part of the section we present calculations of the identified set for an average partial effect.
	
	\subsection{Linear programming representation}
	
	We show that the identified set $\Theta^I$, defined by set \eqref{eq_Theta_I} above, can be represented as a set of $\theta$ values for which a certain linear program has a solution. This characterization facilitates numerical computation of the identified set. 
	
	To present our construction, let us first focus on the $T=2$ case, and suppose that Assumption \ref{ass_disc} holds, so $\alpha_i$ has discrete support. For any hypothetical values $(\widetilde{\theta},\widetilde{\pi},\widetilde{G})\in \Theta\times \Pi\times {\cal{G}}_2$, we define
	\begin{equation}\psi_{x_1}(x_2,y_1,\alpha)=\Pr(X_{i2}=x_2,Y_{i1}=y_1,\alpha_i=\alpha\,|\, X_{i1}=x_1;\widetilde{\theta},\widetilde{\pi},\widetilde{G}).\label{eq_psi}\end{equation} The right-hand-side of \eqref{eq_psi} is determined by the unknown heterogeneity distribution, the parametric likelihood for $Y_1$ (given $X_1$ and $\alpha$), and the unknown feedback process for $X_2$. Finding $\Theta_I$ essentially involves repeatedly asking whether, for a given $\widetilde{\theta}$, there exists a valid feedback process and heterogeneity distributions consistent with the observed data distribution (and the parametric part of the model).
	
	Specifically we first require that $\psi_{x_1}(x_2,y_1,\alpha)$ is a valid probability mass function:
	\begin{equation}\psi_{x_1}(x_2,y_1,\alpha)\geq 0, \quad \sum_{x_2=0}^1\sum_{y_1=0}^1\int _{{\cal{S}}}\psi_{x_1}(x_2,y_1,\alpha)d\mu(\alpha)=1.\label{eq_IS2}\end{equation} Second, we check that it is consistent with the parametric likelihood model for $Y_1$ given $X_1$ and $\alpha$:
	\begin{equation}\sum_{x_2=0}^1\psi_{x_1}(x_2,y_1,\alpha)= F(\widetilde{\theta} x_1+\alpha)^{y_1}[1-F(\widetilde{\theta} x_1+\alpha)]^{1-y_1}\sum_{x_2=0}^1\sum_{y_1=0}^1\psi_{x_1}(x_2,y_1,\alpha).\label{eq_IS3}\end{equation}
	Finally, we conclude that $\widetilde{\theta}\in \Theta^I$ if and only if \begin{equation}Q_{x_1}(y_2,y_1,x_2;\theta,\pi,G)=\int_{{\cal{S}}} F(\widetilde{\theta} x_2+\alpha)^{y_2}[1-F(\widetilde{\theta} x_2+\alpha)]^{1-y_2}\psi_{x_1}(x_2,y_1,\alpha)d\mu(\alpha),\label{eq_IS1}\end{equation}
	for some vectors $\psi_{x_1}$ also satisfying (\ref{eq_IS2}) and (\ref{eq_IS3}) for $x_1\in\{0,1\}$. Condition \eqref{eq_IS1} ensures compatibility with the likelihood contribution for the period $2$ outcome, $Y_2$.
	
	Since all of the equalities and inequalities in (\ref{eq_IS2}), (\ref{eq_IS3}) and (\ref{eq_IS1}) are linear in $\psi_{x_1}$, it follows that one can verify whether $\widetilde{\theta}\in \Theta^I$ by checking the existence of a solution to a finite-dimensional linear program.\footnote{Note that, to compute the identified set under the assumption of strict exogeneity, one can simply modify this approach by adding to (\ref{eq_IS2}), (\ref{eq_IS3}) and (\ref{eq_IS1}) the additional restriction \begin{equation*}\frac{\psi_{x_1}(x_2,1,\alpha)}{F(\widetilde{\theta} x_1+\alpha)}=\frac{\psi_{x_1}(x_2,0,\alpha)}{1-F(\widetilde{\theta} x_1+\alpha)}\quad \text{for all }(x_2,x_1,\alpha),\end{equation*}
		which is also linear in $\psi_{x_1}$. The fact that, under strict exogeneity, $\Theta^I$ can be computed using linear programming was first established by \citet{honore2006bounds}.} We provide details about computation in Appendix \ref{App_implement}.
	
	The characterization of $\Theta^I$ in (\ref{eq_IS2}), (\ref{eq_IS3}) and (\ref{eq_IS1}) remains valid when Assumption \ref{ass_disc} does not hold, and $\alpha_i$ has continuous support. In that case, one needs to interpret $\psi_{x_1}$ in (\ref{eq_psi}) as the product between the density of $\alpha_i$ conditional on $(X_{i2},Y_{i1})$ and the probability of $(X_{i2},Y_{i1})$, both of them conditional on $X_{i1}$ and for hypothetical parameter values. The resulting linear program is infinite-dimensional in that case.

	The linear programming representation of $\Theta^I$ extends to any number $T\geq 2$ of periods. To see this, let, for some $(\widetilde{\theta},\widetilde{\pi},\widetilde{G})\in \Theta\times \Pi\times {\cal{G}}_T$,
	$$\psi_{x_1}(x^{2:T},y^{T-1},\alpha)=\Pr(X_{i}^{2:T}=x^{2:T},Y_{i}^{T-1}=y^{T-1},\alpha_i=\alpha\,|\, X_{i1}=x_1;\widetilde{\theta},\widetilde{\pi},\widetilde{G}),$$ with a similar definition when the support of $\alpha_i$ is not discrete and Assumption \ref{ass_disc} does not hold. In Appendix \ref{App_IS} we derive the following characterization of the (sharp) identified set $\Theta^I$.
	\begin{proposition}\label{lem_thetaI} (\textsc{Identified Set})
		$\widetilde{\theta}\in \Theta^I$ if, and only if,\begin{equation}Q_{x_1}(y^{T},x^{2:T};\theta,\pi,G)=\int_{{\cal{S}}} F(\widetilde{\theta} x_T+\alpha)^{y_T}[1-F(\widetilde{\theta} x_T+\alpha)]^{1-y_T}\psi_{x_1}(x^{2:T},y^{T-1},\alpha)d\mu(\alpha),\label{eq_IS1_T}\end{equation}
		for some integrable functions $\psi_{x_1}:\{0,1\}^{2T-2}\times {\cal{S}}\rightarrow \mathbb{R}$, $x_1\in\{0,1\}$, satisfying
		\begin{equation}\psi_{x_1}(x^{2:T},y^{T-1},\alpha)\geq 0, \quad \sum_{x^{2:T}\in\{0,1\}^{T-1}}\sum_{y^{T-1}\in\{0,1\}^{T-1}}\int_{{\cal{S}}}\psi_{x_1}(x^{2:T},y^{T-1},\alpha)d\mu(\alpha)=1,\label{eq_IS2_T}\end{equation}
		and, for all $s\in\{2,...,T\}$,\footnote{For $s=T$, restriction (\ref{eq_IS3_T}) should be read as 
				\begin{align*}\sum_{x_{T}=0}^1\psi_{x_1}(x^{2:T},y^{T-1},\alpha)
			= F(\widetilde{\theta} x_{T-1}+\alpha)^{y_{T-1}}[1-F(\widetilde{\theta} x_{T-1}+\alpha)]^{1-y_{T-1}}\sum_{x_T=0}^1\sum_{y_{T-1}=0}^1\psi_{x_1}(x^{2:T},y^{T-1},\alpha).	\end{align*} } also satisfying
		\begin{align}&\sum_{x^{s:T}\in\{0,1\}^{T-s+1}}\sum_{y^{s:T-1}\in\{0,1\}^{T-s}}\psi_{x_1}(x^{2:T},y^{T-1},\alpha)\notag\\
			&= F(\widetilde{\theta} x_{s-1}+\alpha)^{y_{s-1}}[1-F(\widetilde{\theta} x_{s-1}+\alpha)]^{1-y_{s-1}}\sum_{x^{s:T}\in\{0,1\}^{T-s+1}}\sum_{y^{s-1:T-1}\in\{0,1\}^{T-s+1}}\psi_{x_1}(x^{2:T},y^{T-1},\alpha).\label{eq_IS3_T}	\end{align} 
	\end{proposition}
	Proposition \ref{lem_thetaI} shows that one can verify whether $\widetilde{\theta}\in \Theta^I$ by checking the feasibility of a (finite- or infinite-dimensional) linear program. In a setting with lagged outcomes and strictly exogenous covariates, \citet{honore2006bounds} provided an analogous linear programming representation of the identified set. By contrast, in Proposition \ref{lem_thetaI} we characterize the identified set of $\theta$ in the general predetermined case where the Granger condition fails; i.e., when $G_{y^{t-1},x^{t-1}}(\alpha)$ may depend on $y^{t-1}$, a situation that \citet{honore2006bounds} did not consider but anticipated in their conclusion.

	\subsection{Numerical illustration}

	In this section we compute identified sets $\Theta^I$ in logit and probit models for a set of example data generating processes (DGPs). In the DGPs, $X_{it}$ follows a Bernoulli distribution on $\{0,1\}$ with probabilities $(\frac{1}{2},\frac{1}{2})$, independent over time, and $\alpha_i$ takes $K=31$ values with probabilities closely resembling those of a standard normal (a specification we borrow from \citealp{honore2006bounds}), and is drawn independently of $(X_{i1},...,X_{iT})$. In the logit case, $F(u)=\frac{e^u}{1+e^u}$, and in the probit case, $F(u)=\Phi(u)$ for $\Phi$ the standard normal cdf. Lastly, we vary $\theta$ between $-1$ and $1$. Note that $X_{it}$ is strictly exogenous in this data generating process. We characterize identified sets in two scenarios: assuming that $X_{it}$ are strictly exogenous, and only assuming that $X_{it}$ are predetermined.

\begin{figure}[tbp]
		\caption{Identified sets in logit and probit models\label{Fig_logit}}
		\begin{center}
			\begin{tabular}{cc}
				LOGIT MODEL & PROBIT MODEL\\
				\multicolumn{2}{c}{$T=2$}   \\
				\includegraphics[width=80mm, height=55mm]{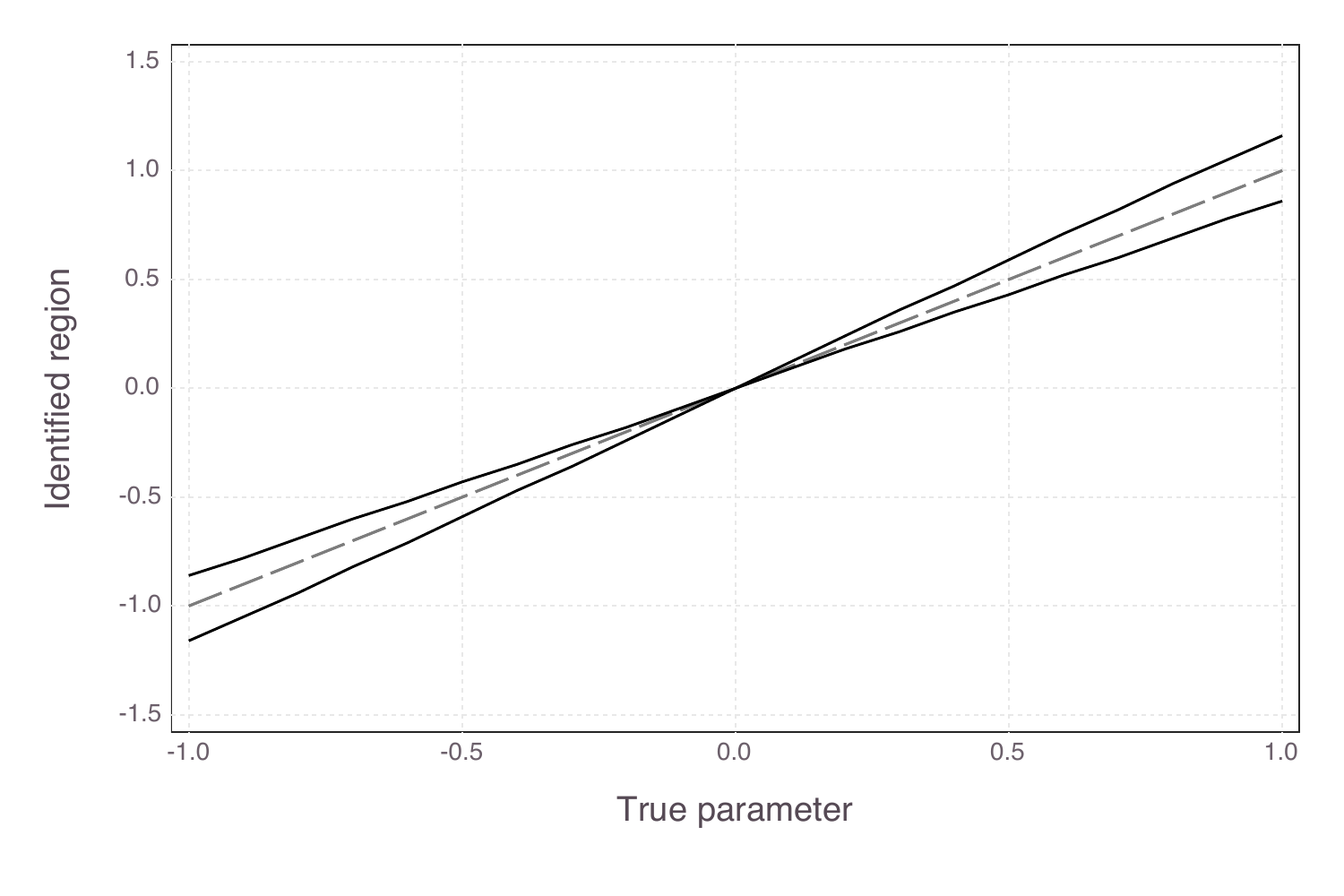} & \includegraphics[width=80mm, height=55mm]{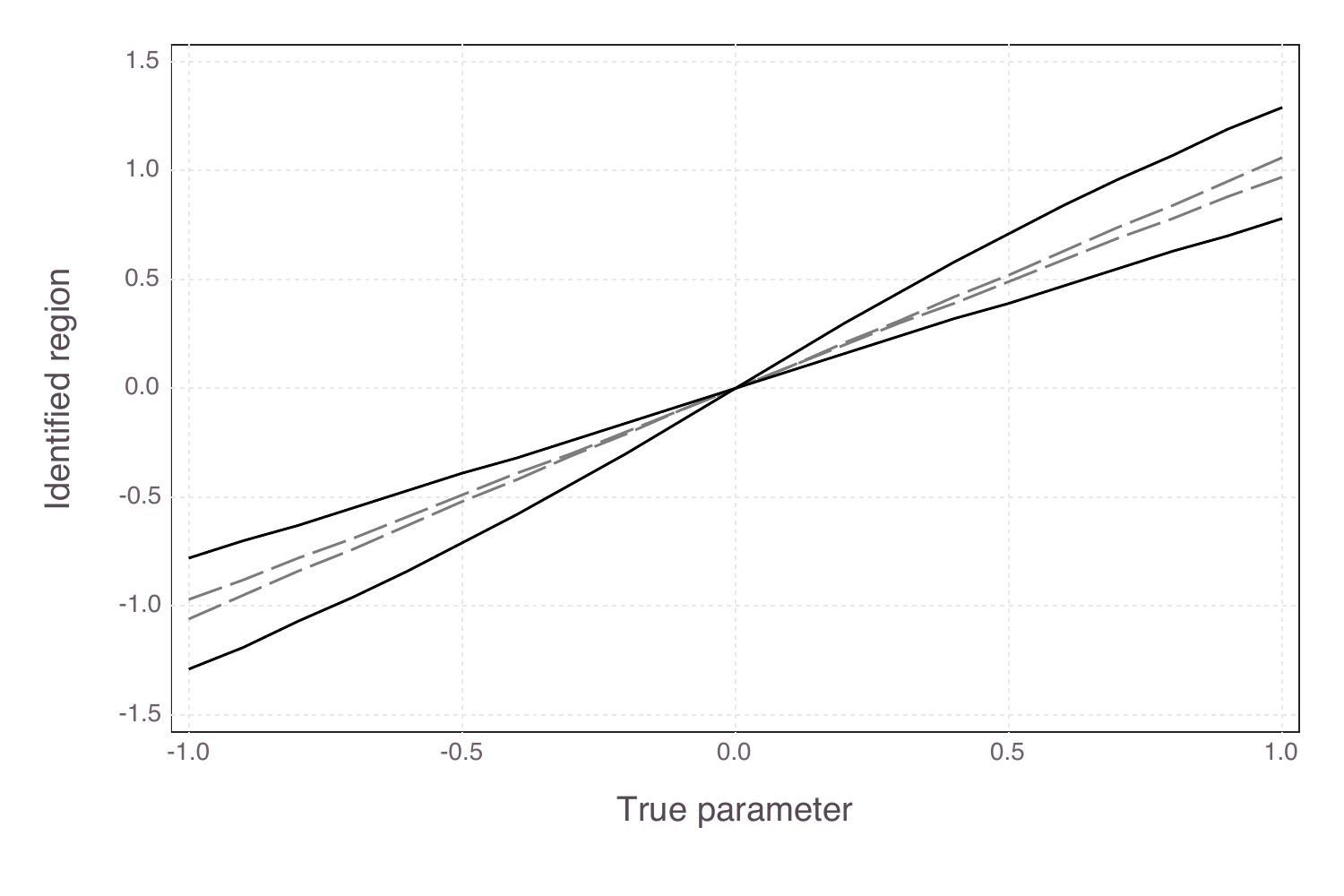}\\
				\multicolumn{2}{c}{$T=3$}\\ \includegraphics[width=80mm, height=55mm]{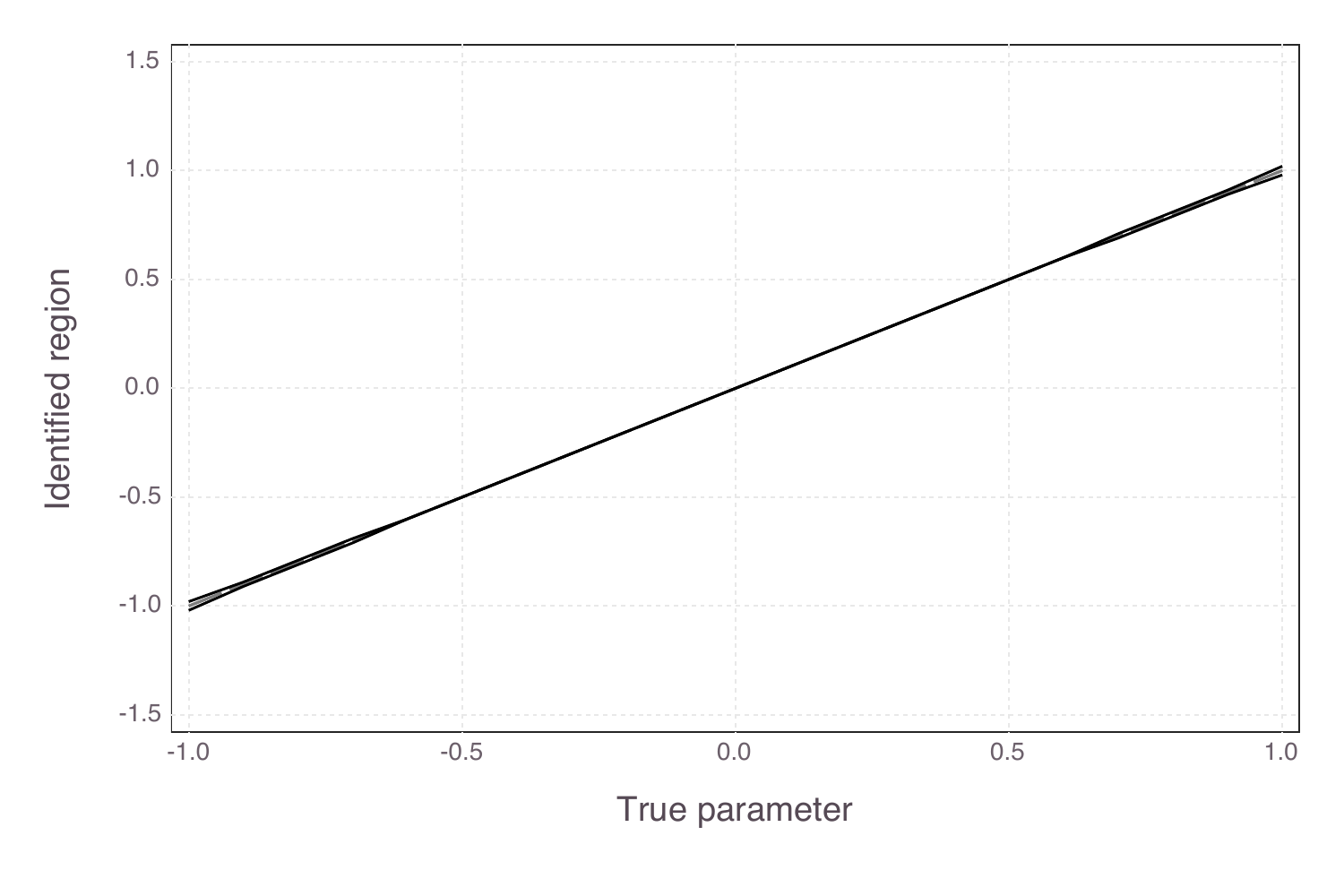} &\includegraphics[width=80mm, height=55mm]{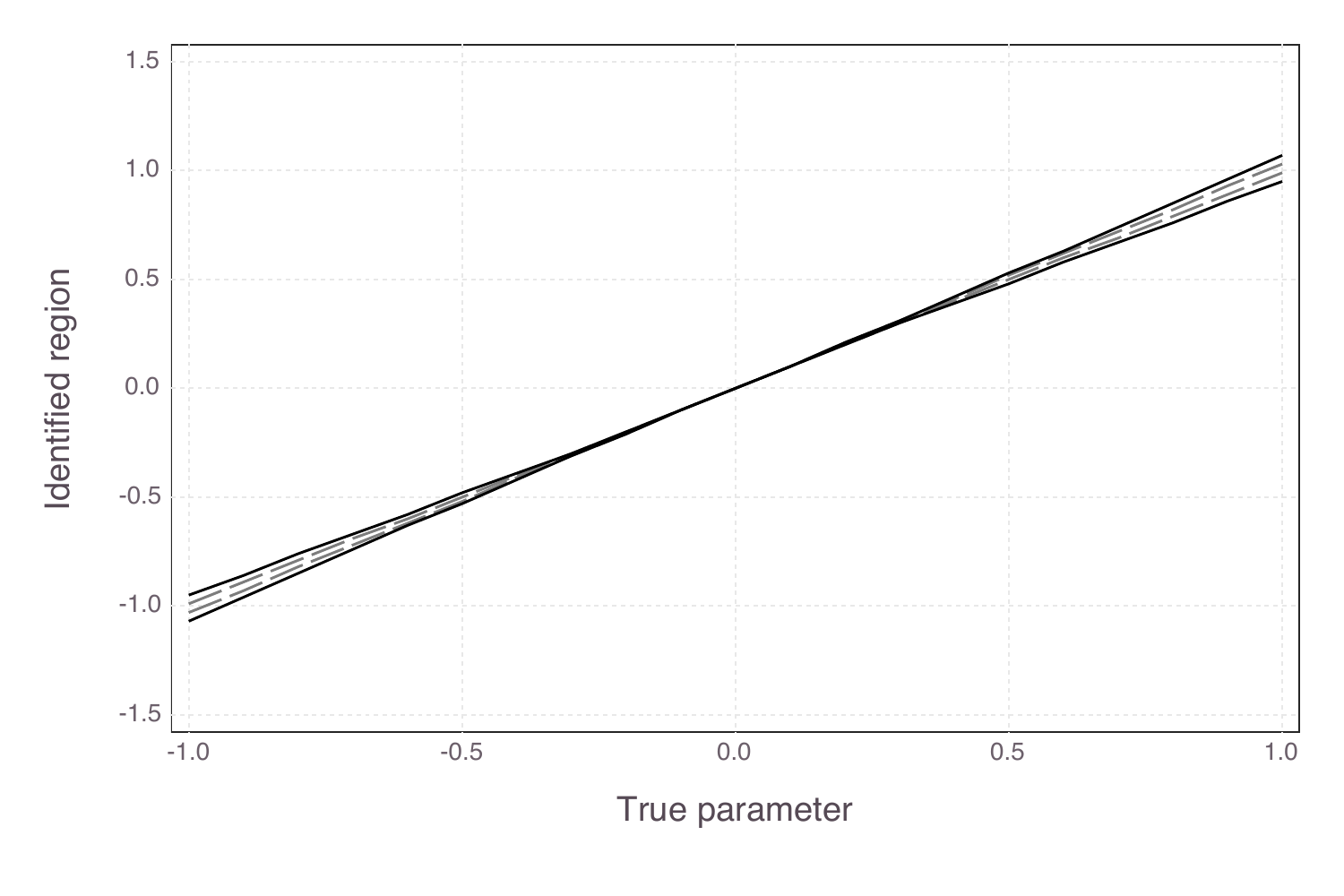}\\
				\multicolumn{2}{c}{$T=4$}\\ \includegraphics[width=80mm, height=55mm]{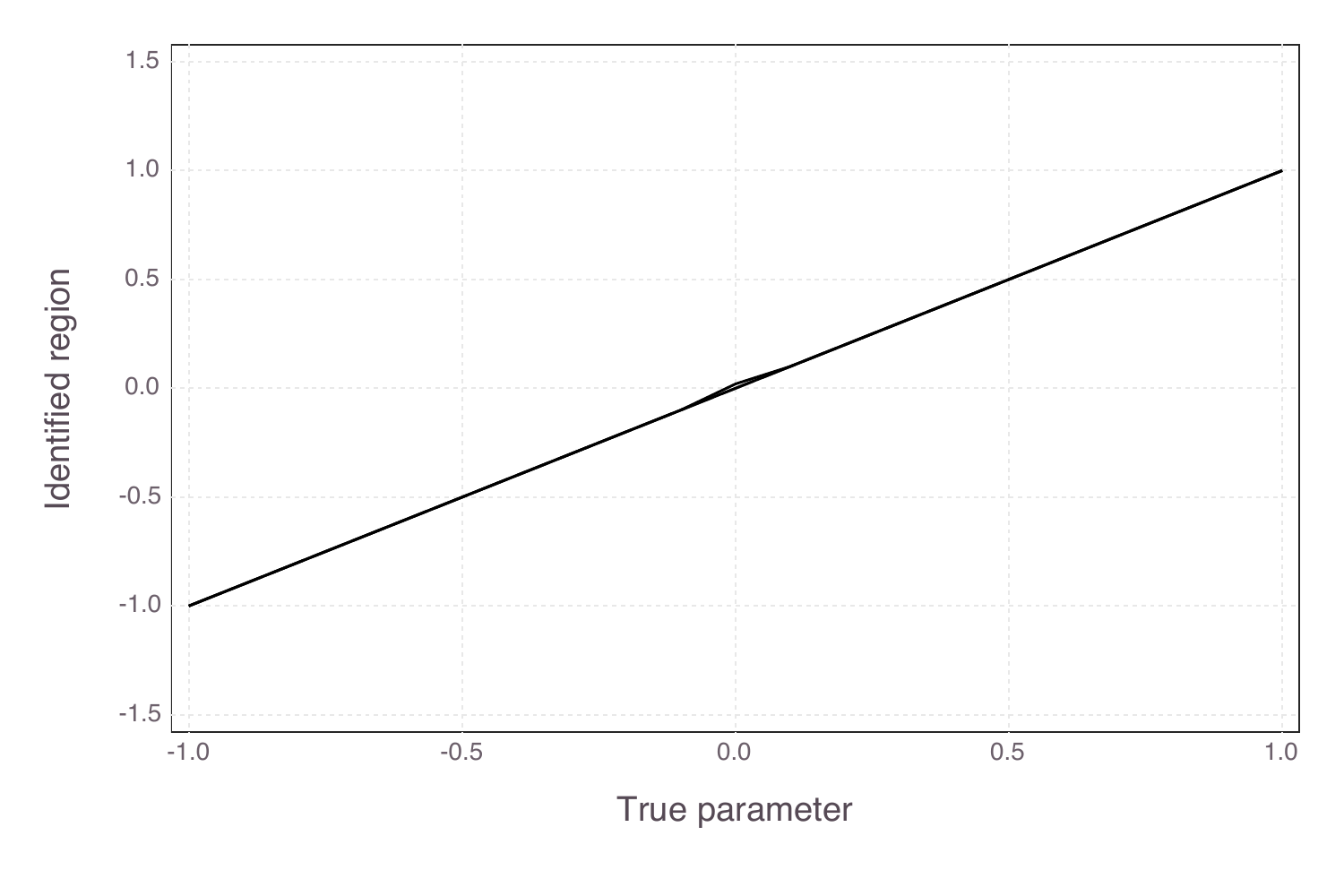}   & \includegraphics[width=80mm, height=55mm]{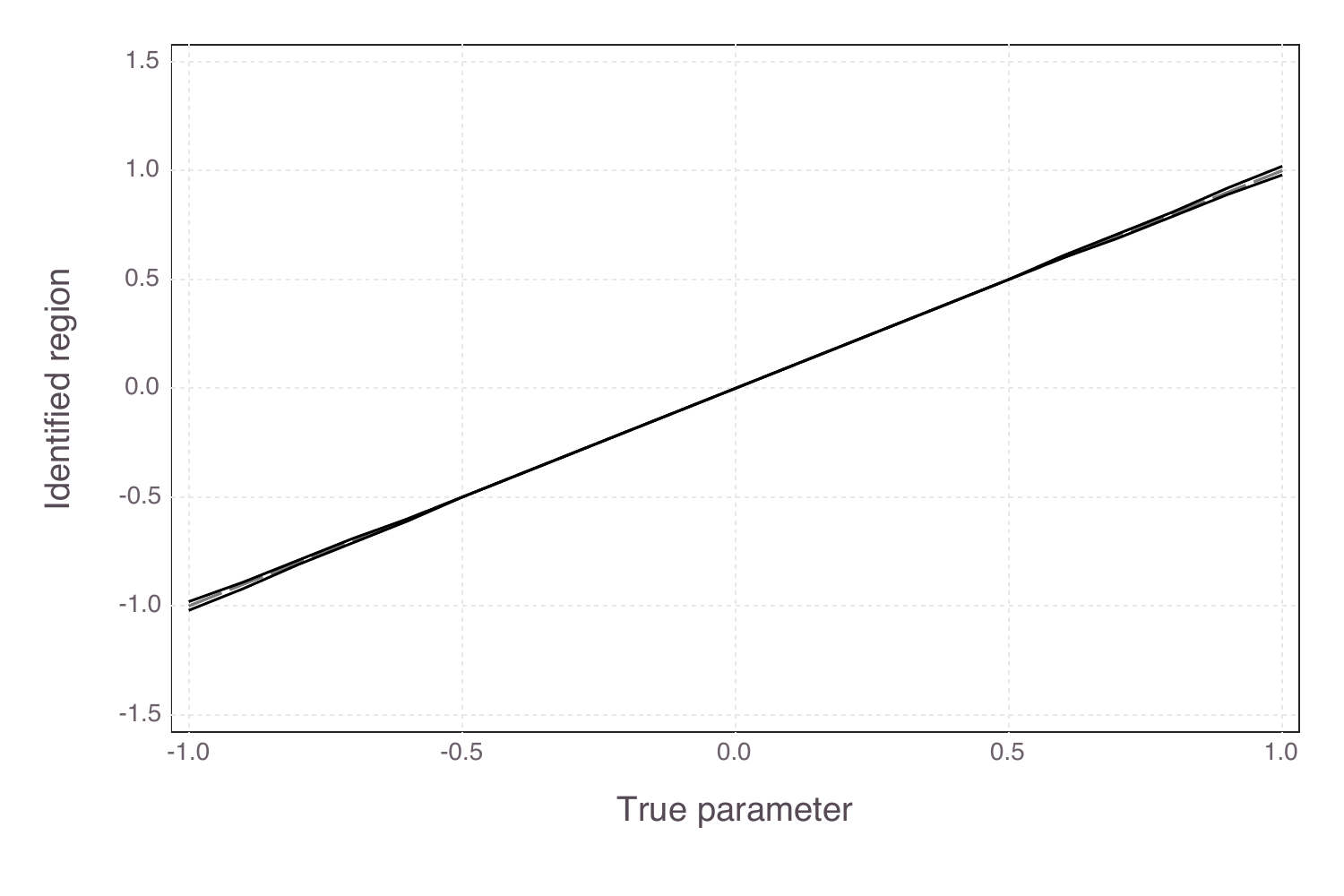}\end{tabular}
		\end{center}
		\par
		\textit{{\footnotesize Notes: Upper and lower bounds of the identified set $\Theta^I$ in a logit model (left column) and a probit model (right column), for $T=2,3,4$. The identified sets under strict exogeneity are indicated by the dashed lines, the sets under predeterminedness are indicated by the solid lines. The population value of $\theta$ is given on the x-axis.}}
	\end{figure}

	In Figure \ref{Fig_logit} we report our numerical calculations of the identified set $\Theta^I$ for the logit model (in the left column panels) and for the probit model (in the right column panels). The three vertical panels correspond to the $T=2,3,4$ cases, respectively. In each graph, we report two sets of upper and lower bounds: those computed while maintaining the strict exogeneity assumption (in dashed lines) and those computed maintaining just predeterminedness (in solid lines). We report the true parameter $\theta$ on the x-axis. To compute the sets, we assume that $\alpha_i$ has the same $K=31$ points of support as in the DGP. We also experimented with fewer and additional support points, as we report below.

	Focusing first on the logit case, shown in the left column of Figure \ref{Fig_logit}, we see that the identified set $\Theta^I$ under strict exogeneity is a singleton for any value of $\theta$ and irrespective of $T$. This is not surprising since $\theta$ is point-identified in the static logit model. In contrast, the upper and lower bounds of the identified set do not coincide in the predetermined case, consistent with our non point-identification result. At the same time, the identified sets appear rather narrow, even when $T=2$, and the width of the set tends to decrease rapidly when $T$ increases to three and four periods. This is qualitatively similar to the observation of \citet{honore2006bounds}, who focused on dynamic probit models and found that the width of the identified set tends to decrease rapidly with $T$.
	
	
	Focusing next on the probit case, shown in the right column of Figure \ref{Fig_logit}, we see that the identified set $\Theta^I$ under strict exogeneity is not a singleton. Moreover, allowing the covariate to be predetermined increases the width of the identified set. However, as in the logit case, the sets appear rather narrow, even when $T=2$, and their widths decrease quickly as $T$ increases. Of course, these observations are specific to a particular data-generating process and the corresponding bounds may be wide for other DGPs.
	
	The results in Figure \ref{Fig_logit} are obtained by assuming that the researcher knows the (finite) support of $\alpha_i$. This approach is similar to the one in \citet{honore2006bounds}. Alternatively, one may wish to characterize the identified set in a class of models where $\alpha_i$ is continuous, e.g., when ${\cal{S}}=\mathbb{R} $ and $\mu$ is the Lebesgue measure. Doing so, as noted earlier, requires approximating an infinite-dimensional linear program.   In Appendix Figure \ref{App_Fig_IS_rob}, we go take a heuristic step in this direction by reporting numerical approximations to the identified sets, for $T=2$, obtained by taking $K=5$, $K=50$, and $K=500$ points of support for $\alpha_i$, respectively, where the points of support are equidistant percentiles of a standard normal distribution. We find very minor differences compared to the case $K=31$ that we report in Figure \ref{Fig_logit}. While we do not provide a formal analysis of numerical approximation properties, this suggests that identified sets under continuous $\alpha_i$ may not be markedly different from the ones in Figure \ref{Fig_logit}. 
	
	Overall, these calculations suggest that, while relaxing strict exogeneity tends to increase the widths of the bounds, the identified sets under predeterminedness can be informative even when the number of periods is very small. To reiterate, these conclusions are based on a particular set of example DGPs.

	\subsection{Average partial effect}
	
	Although our focus in this paper is on the parameter $\theta$, in applications researchers are often interested in average partial effects such as
	\begin{equation}
		\Delta=\mathbb{E}[\Pr(Y_{it}=1\,|\, X_{it}=1,\alpha_i)-\Pr(Y_{it}=1\,|\, X_{it}=0,\alpha_i)],\label{eq_APE}
	\end{equation}
	where the expectation is taken with respect to the distribution of $\alpha_i$.
	
	The identified set for $\Delta$ can also be characterized as the solution to a linear program. Indeed, it follows from Proposition \ref{lem_thetaI} that $\widetilde{\Delta}$ is in the identified set of $\Delta$ if and only if there exists $\widetilde{\theta}$, $\psi_{0}$ and $\psi_1$ such that (\ref{eq_IS1_T}), (\ref{eq_IS2_T}), and (\ref{eq_IS3_T}) hold, and 
	\begin{equation}\label{eq_mu_IS}
		\widetilde{\Delta}= \int_{\cal{S}}[F(\widetilde{\theta}+\alpha)-F(\alpha)]\sum_{x_1\in\{0,1\}}q_{x_1}\sum_{x^{2:T}\in\{0,1\}^{T-1}}\sum_{y^{T-1}\in\{0,1\}^{T-1}}\psi_{x_1}(x^{2:T},y^{T-1},\alpha)d\mu(\alpha), 
	\end{equation}
	where $q_{x_1}=\Pr(X_{i1}=x_1)$. For any given $\widetilde{\theta} \in\Theta^I$, we can therefore compute the set of $\widetilde{\Delta}$ parameters in the identified set by solving a linear program. We provide details about computation in Appendix \ref{App_implement}. 
	
	\begin{figure}[tbp]
		\caption{Identified sets for average partial effects in logit and probit models\label{Fig_APE_logit}}
		\begin{center}
			\begin{tabular}{cc}
				LOGIT MODEL & PROBIT MODEL\\
				\multicolumn{2}{c}{$T=2$}   \\
				\includegraphics[width=80mm, height=55mm]{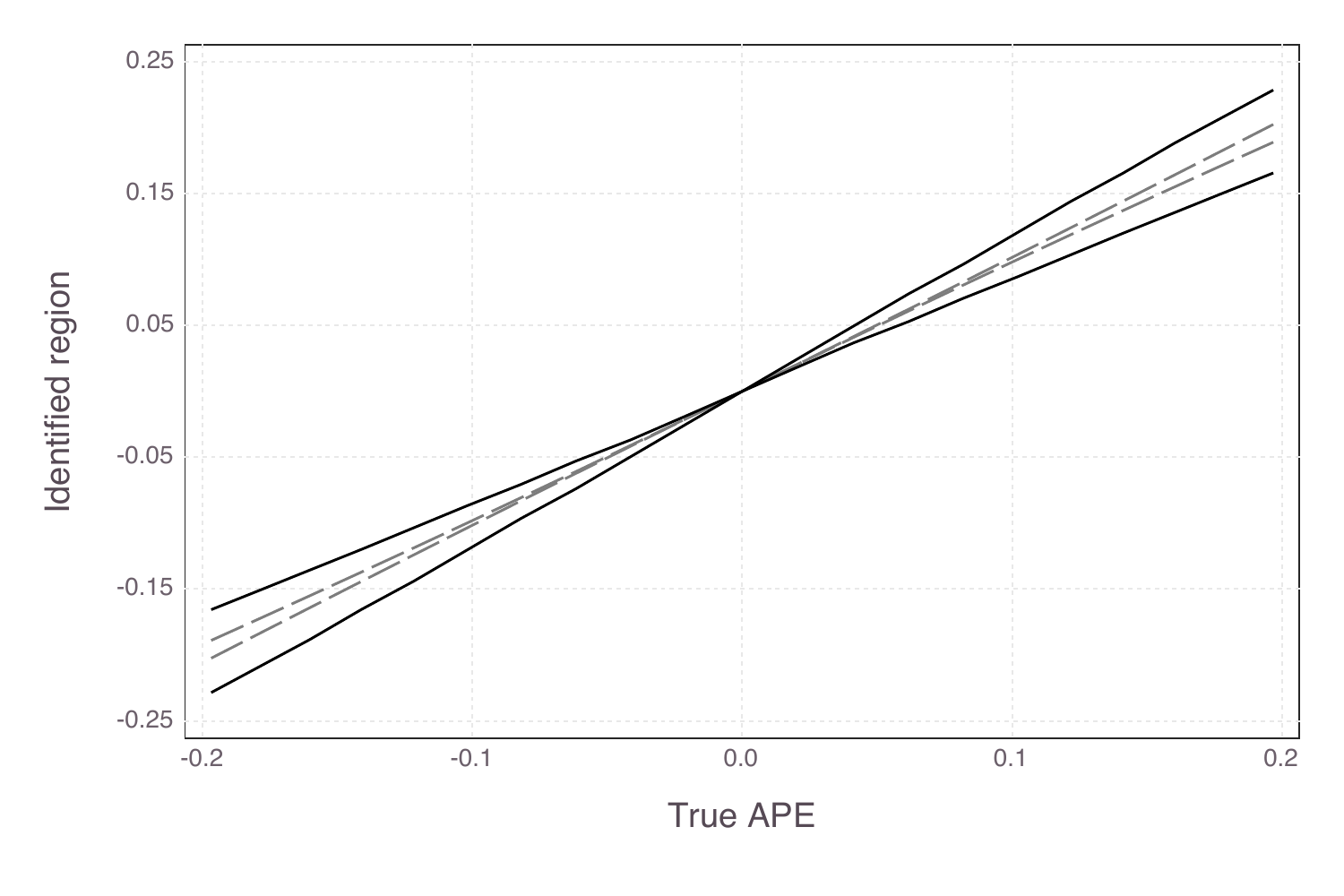} & \includegraphics[width=80mm, height=55mm]{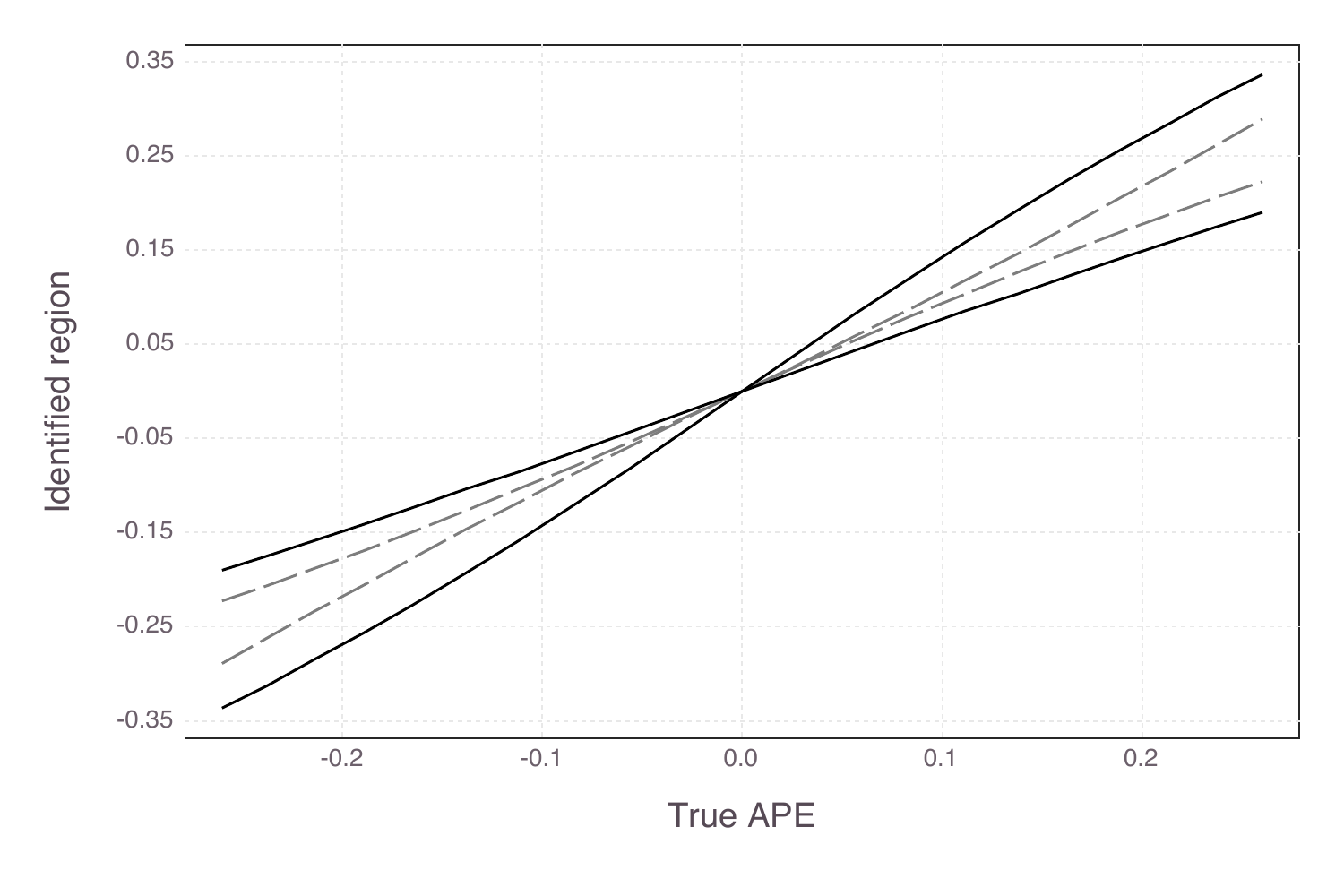}\\
				\multicolumn{2}{c}{$T=3$}\\ \includegraphics[width=80mm, height=55mm]{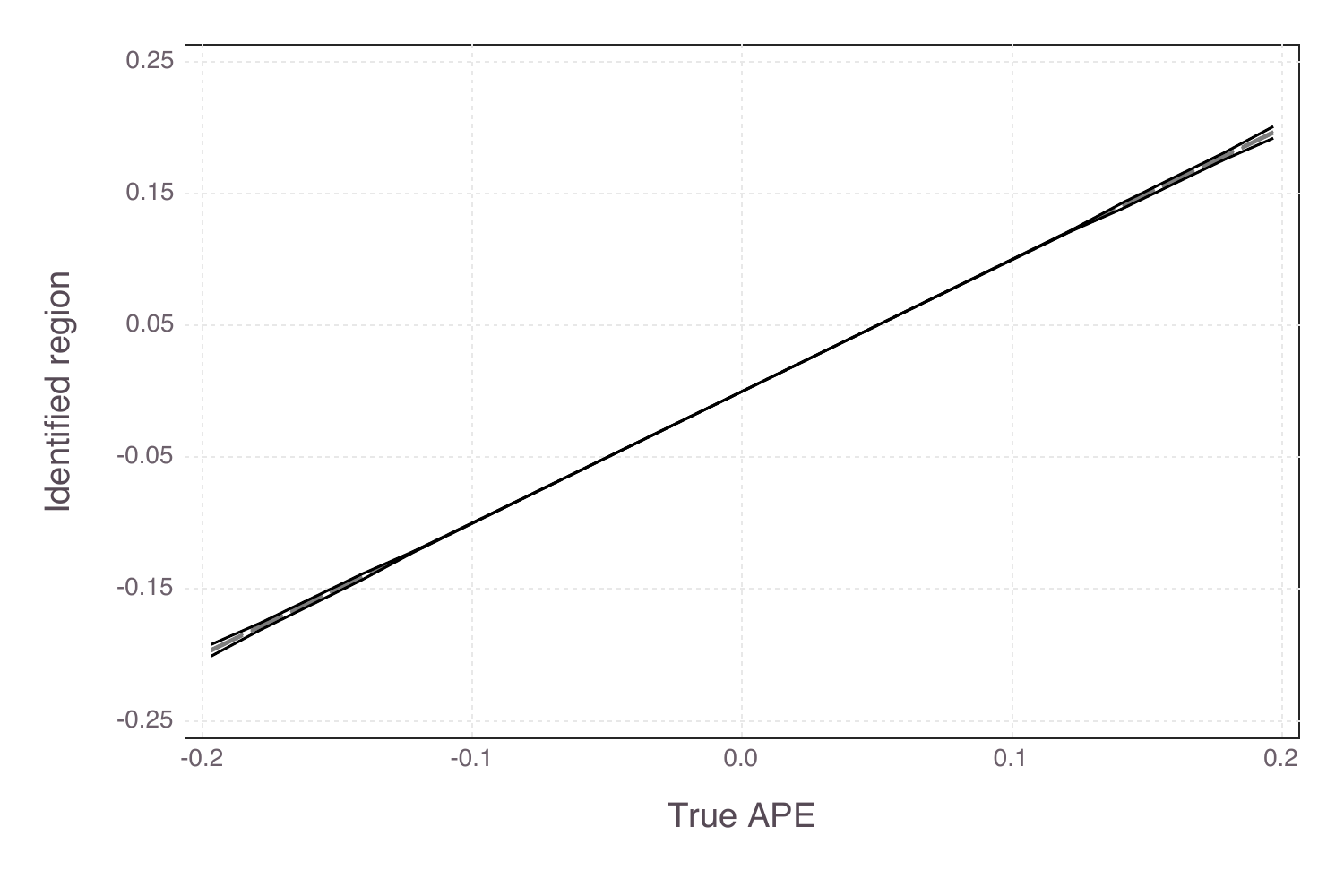} &\includegraphics[width=80mm, height=55mm]{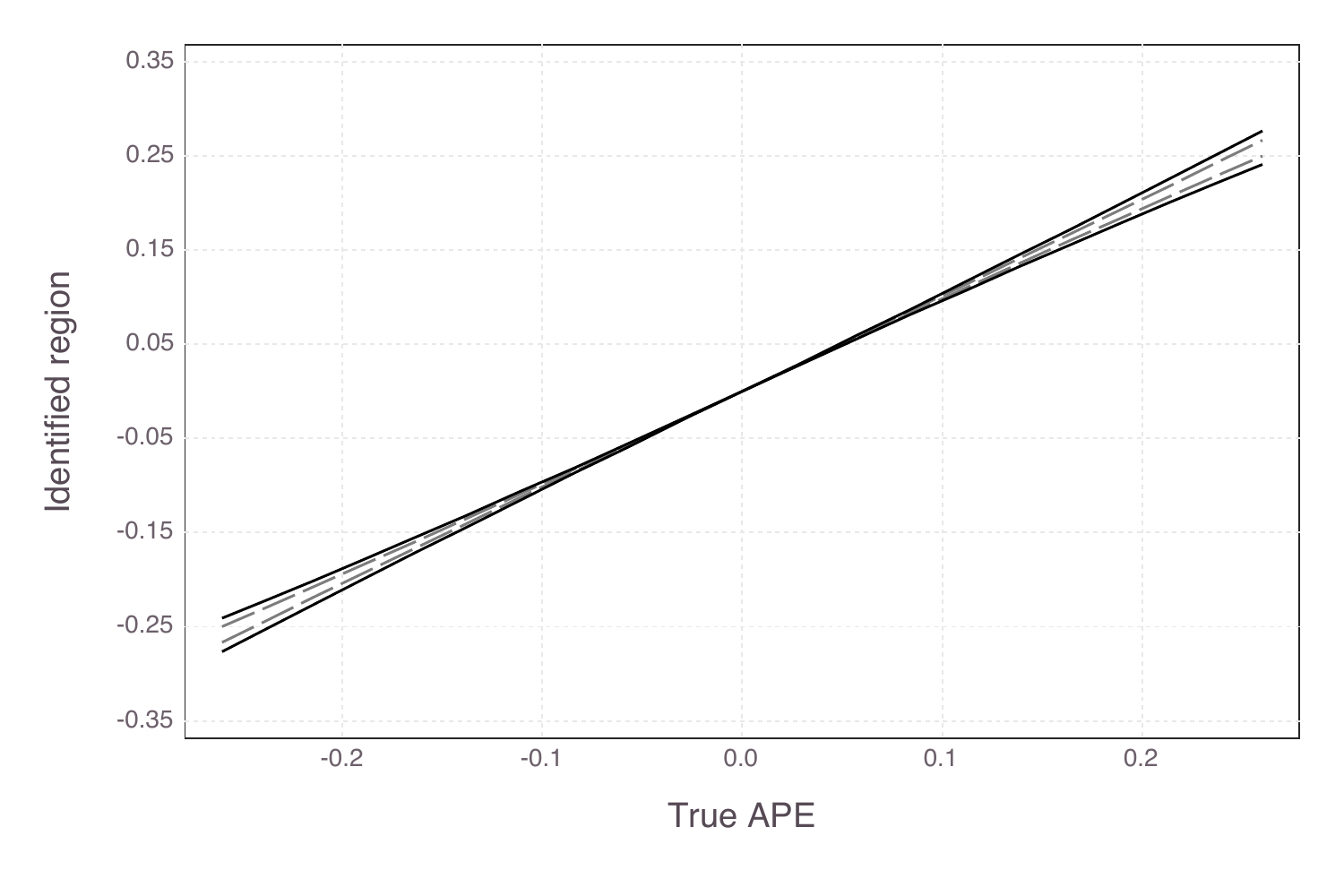}\\
				\multicolumn{2}{c}{$T=4$}\\ \includegraphics[width=80mm, height=55mm]{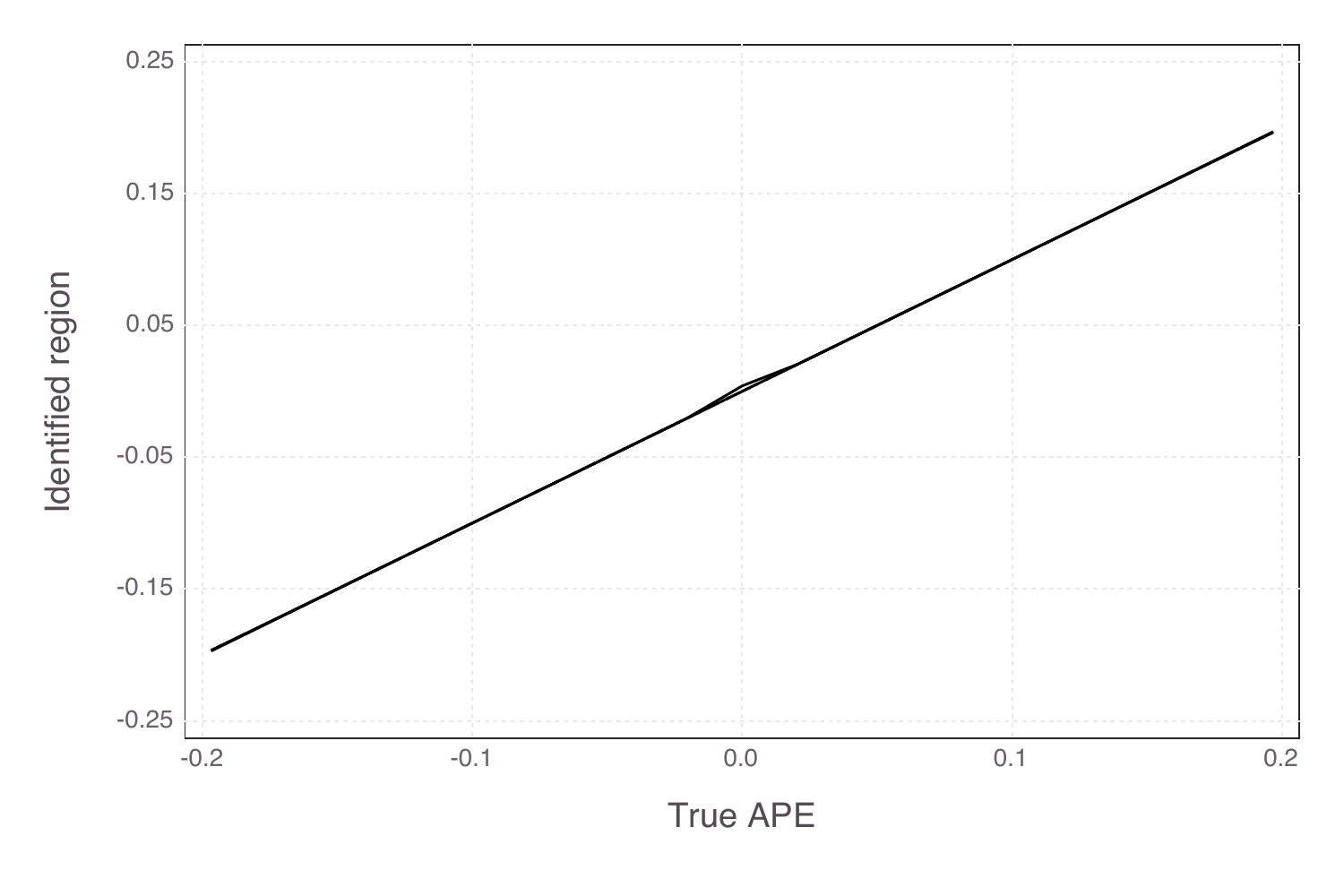}   & \includegraphics[width=80mm, height=55mm]{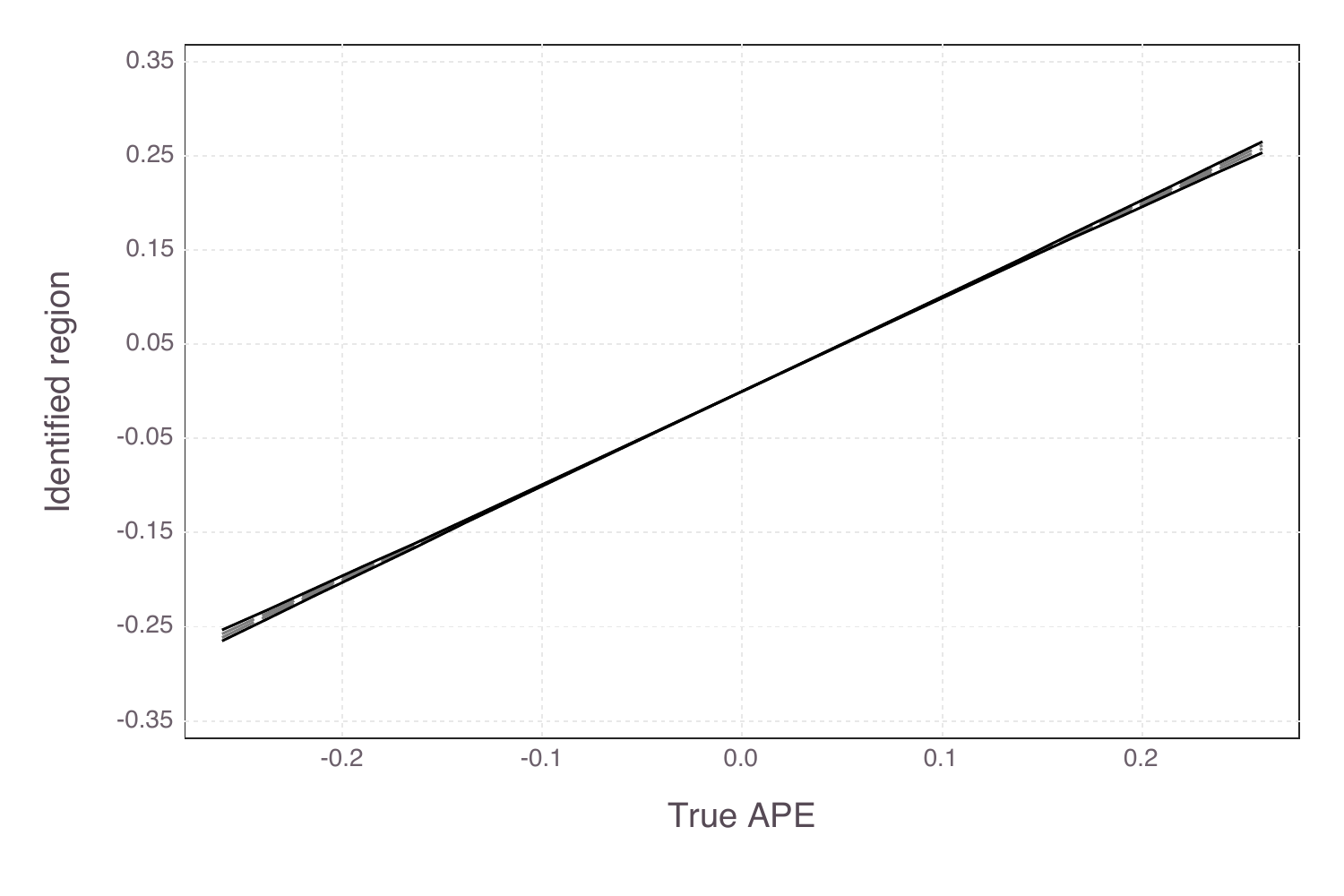}
			\end{tabular}
		\end{center}
		\par
		\textit{{\footnotesize Notes: Upper and lower bounds of the identified set for the average partial effect in a logit model (left column) and a probit model (right column), for $T=2,3,4$. The identified sets under strict exogeneity are indicated by the dashed lines, the sets under predeterminedness are indicated by the solid lines. The population value of the average partial effect is given on the x-axis.}}
	\end{figure}

	In Figure \ref{Fig_APE_logit} we report our computations of the identified set for the average partial effect $\Delta$, relying on the same parameter values and DGP as before. Focusing first on the logit case, shown in the left column of the figure, we see that the identified set under strict exogeneity is not a singleton, except when the true $\theta$ and $\Delta$ are equal to zero. This is not surprising, since average partial effects  generally fail to be point-identified in binary choice models, even when covariates are strictly exogenous. Yet, the sets seem rather narrow, even when $T=2$. Allowing the covariate to be predetermined increases the widths of the sets, however the increase is relatively moderate. Moreover, the sets under predeterminedness are very tight whenever $T\geq 3$.

	Focusing next on the probit case, shown in the right column of Figure \ref{Fig_APE_logit}, we see that although the sets appear wider than in the logit case, relaxing strict exogeneity only moderately increases the widths of the sets, especially when $T\geq 3$.

	Lastly, while we compute the sets in Figure \ref{Fig_APE_logit} under the assumption that $\alpha_i$ has the same $K=31$ points of support as in the DGP, in Appendix Figure \ref{App_Fig_IS_APE_rob} we report approximations of the sets, for $T=2$, obtained using $K=5$, $K=50$, and $K=500$ points of support for $\alpha_i$. The sets appear very similar to the ones based on $K=31$ points of support shown in Figure \ref{Fig_APE_logit}. However, in this case as well, we do not formally analyze the numerical approximation of the identified sets under continuous $\alpha_i$.

	\section{Restrictions on the feedback process\label{sec_restrict}}
	
	Our analysis suggests that failures of point-identification are commonplace in binary choice models with a predetermined covariate. In this section we describe possible restrictions on the model that can strengthen its identification content. We focus on restrictions on the feedback process, since restrictions on individual heterogeneity are rarely motivated by the economic context.     
	
	%
	%
	%
	%
	%
	%
	%
	
	\subsection{Homogeneous feedback}
	
	In some applications one may want to restrict the feedback process to not depend on time-invariant heterogeneity; that is, to impose that
	\begin{equation}\label{feedback_homog}\Pr\left(X_{it}=1\,|\, Y_{i}^{t-1}=y^{t-1},X_{i}^{t-1}=x^{t-1},\alpha_i=\alpha\right)=G^t_{y^{t-1},x^{t-1}}\end{equation}
	is independent of $\alpha$. For example, in structural dynamic discrete choice models, researchers may be willing to model the law of motion of state variables such as dynamic production inputs as homogeneous across units. \citet{kasahara2009nonparametric} show how this assumption can help identification in these models. Here we study how a homogeneity assumption can lead to tighter identified sets in our setting.
	
	To proceed, we focus on the case where $T=2$. Given (\ref{feedback_homog}), the likelihood function takes the form
	\begin{align*}&\Pr\left(Y_{i2}=y_2,X_{i2}=x_{2},Y_{i1}=y_1\,|\, X_{i1}=x_1\right)\\&=\left\{\int_{{\cal{S}}}\,  F(\theta x_2+\alpha)^{y_2}[1-F(\theta x_2+\alpha)]^{1-y_2}F(\theta x_1+\alpha)^{y_1}[1-F(\theta x_1+\alpha)]^{1-y_1}\pi_{x_1}(\alpha)d\mu(\alpha)\right\}\\
		&\quad\quad \quad \times [G^2_{y_{1},x_1}]^{x_2}[1-G^2_{y_{1},x_{1}}]^{1-x_2},\end{align*}
	where the likelihood factors due to the fact that the feedback process does not depend on $\alpha$. Hence, under Assumption \ref{ass_1} (which avoids division by zero) we have
	\begin{align}&\frac{\Pr\left(Y_{i2}=y_2,X_{i2}=x_{2},Y_{i1}=y_1\,|\, X_{i1}=x_1\right)}{[G^2_{y_{1},x_1}]^{x_2}[1-G^2_{y_{1},x_{1}}]^{1-x_2}}\notag\\&=\int_{{\cal{S}}}\,  F(\theta x_2+\alpha)^{y_2}[1-F(\theta x_2+\alpha)]^{1-y_2}F(\theta x_1+\alpha)^{y_1}[1-F(\theta x_1+\alpha)]^{1-y_1}\pi_{x_1}(\alpha)d\mu(\alpha).\label{eq_predet_homog}\end{align}
	
	A key observation to make about \eqref{eq_predet_homog} is its right-hand-side coincides with the likelihood function of a binary choice model with a strictly exogenous covariate (where in addition $\alpha_i$ is independent of $X_{i2}$ given $X_{i1}$). In turn, the left-hand side is weighted by the inverse of the feedback process. This is similar to the inverse-probability-of-treatment-weighting approach to dynamic treatment effect analysis in Jamie Robins' work (e.g., \citealp{robins2000marginal}), with the difference that here we focus on panel data models with fixed effects.

	The similarity between (\ref{eq_predet_homog}) and the strictly exogenous case directly delivers point-identification results and consistent estimators. For example, suppose that $F$ is logistic. Given that the left-hand side of (\ref{eq_predet_homog}) is point-identified, it follows from standard arguments (\citealp{rasch1960studies}, \citealp{andersen1970asymptotic}) that $\theta$ is point-identified. Moreover, a consistent estimator of $\theta$ is obtained by maximizing the weighted conditional logit log-likelihood 
	$$\sum_{i=1}^n\widehat{\omega}_i\boldsymbol{1}\{Y_{i1}+Y_{i2}=1\}\left\{ Y_{i1}\ln\left(\frac{\exp(\widetilde{\theta}X_{i1} )}{\exp(\widetilde{\theta}X_{i1} )+\exp(\widetilde{\theta}X_{i2} )}\right)+Y_{i2}\ln\left(\frac{\exp(\widetilde{\theta}X_{i2} )}{\exp(\widetilde{\theta}X_{i1} )+\exp(\widetilde{\theta}X_{i2} )}\right)\right\},$$with weights $$\widehat{\omega}_i=\left\{[\widehat{G}^2_{Y_{i1},X_{i1}}]^{X_{i2}}[1-\widehat{G}^2_{Y_{i1},X_{i1}}]^{1-X_{i2}}\right\}^{-1},$$for $\widehat{G}^2_{y_{1},x_1}$ a consistent estimate of the homogeneous feedback probabilities.\footnote{The analysis in this subsection is not restricted to the binary covariate case. However, when $X_{it}$ are continuous, demonstrating $\sqrt{n}$ consistency of $\widehat{\theta}$ would generally require imposing rate-of-convergence and other requirements on the first-step estimation of the $\widehat{\omega}_i$ weights.}

	\subsection{Markovian feedback}
	
	Another possible restriction on the feedback process is a Markovian condition, such as 
	\begin{equation}\label{feedback_markov}\Pr\left(X_{it}=1\,|\, Y_{i}^{t-1}=y^{t-1},X_{i}^{t-1}=x^{t-1},\alpha_i=\alpha\right)=G^t_{y_{t-1},x_{t-1}}(\alpha)\end{equation}
	is independent of $(y^{t-2},x^{t-2})$. Such a condition may be natural in models where $X_{it}$ is the state variable in the agent's economic problem (as in \citealp{rust1987optimal} and \citealp{kasahara2009nonparametric}, for example).
	
	In order to characterize the identified set $\Theta^I$ with the Markovian condition (\ref{feedback_markov}) added, we augment the restrictions (\ref{eq_IS1_T}), (\ref{eq_IS2_T}) and (\ref{eq_IS3_T}) with the fact that, for all $s\in\{2,...,T\}$,
	\begin{align*}&\frac{\sum_{x^{s+1:T}\in\{0,1\}^{T-s+1}}\sum_{y^{s:T-1}\in\{0,1\}^{T-s}}\psi_{x_1}(x^{2:T},y^{T-1},\alpha)}{\sum_{x^{s:T}\in\{0,1\}^{T-s+1}}\sum_{y^{s:T-1}\in\{0,1\}^{T-s}}\psi_{x_1}(x^{2:T},y^{T-1},\alpha)}\end{align*}
	does not depend on $(y^{s-2},x^{s-2})$.\footnote{When $s=T$, this requires that $\frac{\psi_{x_1}(x^{2:T},y^{T-1},\alpha)}{\sum_{x_T=0}^1\psi_{x_1}(x^{2:T},y^{T-1},\alpha)}$ does not depend on $(y^{T-2},x^{T-2})$.}
	
	A difficulty arises in this case since this additional set of restrictions is not linear in $\psi_{x_1}$. As a result, one would need to use different techniques to characterize the identified set in the spirit of Proposition \ref{lem_thetaI}, and to establish conditions for (the failure of) point-identification in the spirit of Corollary \ref{coro_3}. Given this, we leave the analysis of identification in models with Markovian feedback processes to future work.

	\section{Conclusion\label{sec_conc}}
	
	In this paper we study a binary choice model with a binary predetermined covariate. We find that failures of point-identification are widespread in this setting. Point-identification fails in many binary choice models, with apparently only a few exceptions (such as the exponential model). At the same time, our numerical calculations of identified sets suggest that the bounds on the parameter can be narrow, even in very short panels. This suggests that, while the strict exogeneity assumption has identifying content, models with predetermined covariates and feedback may still lead to informative empirical conclusions, both for the coefficients of the covariates and for average partial effects.  
	
	Our analysis of models with a binary covariates can easily be extended to handle general discrete covariates with finite support. In particular, for $\theta$ to be regularly point-identified there need to exist $x_1\neq x_2$ in the support of $X_{it}$ such that $1$, $F(\theta'x_1+\alpha)$, and $F(\theta'x_2+\alpha)$, for $\alpha\in{\cal{S}}$, are linearly dependent. This condition fails in many popular specifications such as the logit. In turn, when $X_{it}$ has finite, non-binary support, the identified set can still be computed as a solution to a linear program, analogously to Proposition \ref{lem_thetaI}. However, the extension to continuous covariates is not straightforward in our setting, in particular since the notion of regularity maintained by Assumption \ref{ass_2} no longer applies.

	Finally, although we have analyzed a binary choice model, our techniques can be used to study other models with stronger identification content, such as models for count data (e.g., Poisson regression, \citealp{wooldridge1997multiplicative}, \citealp{blundell2002individual}) and models with continuous outcomes (e.g., censored regression, \citealp{honore2004estimation}, and duration models, \citealp{chamberlain1985heterogeneity}). Deriving sequential moment restrictions in such nonlinear models was considered by \cite{Chamberlain_JOE2022} and is further explored in our companion paper (\citealp{BDG_WP}).

	\vskip 3cm
	
	{\small
		\bibliography{Draft_SERIES_revision}
	}
	\clearpage
	
	\appendix
	\renewcommand{\thesection}{\Alph{section}}

	\setcounter{figure}{0}\renewcommand{\thefigure}{\arabic{figure}}
	
	\setcounter{table}{0}\renewcommand{\thetable}{\arabic{table}}
	
	\setcounter{footnote}{0}\renewcommand{\thefootnote}{\arabic{footnote}}

	\setcounter{assumption}{0}\renewcommand{\theassumption}{A\arabic{assumption}}
	
	\setcounter{equation}{0}\renewcommand{\theequation}{A\arabic{equation}}
	
	\setcounter{lemma}{0}\renewcommand{\thelemma}{A\arabic{lemma}}
	
	\setcounter{proposition}{0}\renewcommand{\theproposition}{A\arabic{proposition}}
	
	\setcounter{corollary}{0}\renewcommand{\thecorollary}{A\arabic{corollary}}
	
	\setcounter{theorem}{0}\renewcommand{\thetheorem}{A\arabic{theorem}}
	
	\hypersetup{colorlinks=true,linkcolor=blue,urlcolor=blue,citecolor=blue}

	\vskip 3cm
	\begin{center}
		{ {\LARGE APPENDIX} }
	\end{center}
	
	\section{Proof of Lemma \ref{lem_1}}

	For any $m\times n$ matrix $A$, we will denote as $${\cal{R}}(A)=\{Au\,:\, u\in \mathbb{R}^n\}$$ the range of $A$, $${\cal{N}}(A)=\{u\in \mathbb{R}^n\,:\,Au=0 \}$$ the null space of $A$, and $A^{\dagger}$ the Moore-Penrose generalized inverse of $A$. \\
	
	\noindent We now proceed to prove Lemma  \ref{lem_1}.  Since $\theta$ is point-identified, it is locally point-identified. Since $(\theta,\pi,G)$ is a regular point of $\nabla Q(\theta,\pi,G)$ by Assumption \ref{ass_2}, it follows from Theorem 8 in \cite{bekker2001identification} that 
	\begin{align} \label{eq_notinrange}
		\nabla_{\theta'} Q 
		\notin {\cal{R}}\left(
		\begin{bmatrix}
			\nabla_{\pi_{1}'} Q_{1} &\nabla_{G_{1}'} Q_{1} & 0 &  0 \\
			0 & 0 &  \nabla_{\pi_{0}'} Q_{0} & \nabla_{G_{0}'} Q_{0} \\
		\end{bmatrix}\right).
	\end{align}
	
	\noindent Therefore, there must exist $x_1\in\{0,1\}$ such that
	\begin{align}  \label{eq_notinrange_x1}
		\nabla_{\theta'} Q_{x_1}
		\notin {\cal{R}}\left(
		\begin{bmatrix}
			\nabla_{\pi_{x_1}'} Q_{x_1} &\nabla_{G_{x_1}'} Q_{x_1}
		\end{bmatrix}\right),
	\end{align}
	and in the rest of the proof we will fix this $x_1$ value. 
	
	\noindent Let $\widetilde{\phi}_{x_1}$ denote the projection of $\nabla_{\theta'} Q_{x_1}$ onto the orthogonal complement of the vector space spanned by the columns of 
	$\begin{bmatrix}
		\nabla_{\pi_{x_1}'} Q_{x_1} &\nabla_{G_{x_1}'} Q_{x_1}
	\end{bmatrix}$; that is,
	\begin{align*}
		\widetilde{\phi}_{x_1}=\nabla_{\theta'} Q_{x_1}-\begin{bmatrix}
			\nabla_{\pi_{x_1}'} Q_{x_1} &\nabla_{G_{x_1}'} Q_{x_1}
		\end{bmatrix}\begin{bmatrix}
			\nabla_{\pi_{x_1}'} Q_{x_1} &\nabla_{G_{x_1}'} Q_{x_1}
		\end{bmatrix}^{\dagger}\nabla_{\theta'} Q_{x_1}.
	\end{align*}

	\noindent It follows from (\ref{eq_notinrange_x1}) that $\widetilde{\phi}_{x_1}\neq 0$.  Moreover, since $\iota'Q_{x_1}(\theta,\pi,G)=1$, where $\iota$ denotes a conformable vector of ones, we have 
	\begin{align}
		\iota'\nabla_{\theta'} Q_{x_1}=0,\,\,\, \iota'\nabla_{\pi_{x_1}'} Q_{x_1}=0, \,\,\, \iota'\nabla_{G_{x_1}'} Q_{x_1}=0. \label{eq_Q_sumto1}
	\end{align}
	It follows that $\iota'\widetilde{\phi}_{x_1}=0$, implying that $\widetilde{\phi}_{x_1}$ cannot be constant.
	
	\noindent Now, since $v'\widetilde{\phi}_{x_1}=0$ for all $v\in {\cal{R}}\left(
	\begin{bmatrix}
		\nabla_{\pi_{x_1}'} Q_{x_1} &\nabla_{G_{x_1}'} Q_{x_1}
	\end{bmatrix}\right)$,
	we have
	$$\widetilde{\phi}_{x_1} \in {\cal{N}}(\nabla_{\pi_{x_1}} Q_{x_1}')\cap {\cal{N}}(\nabla_{G_{x_1}} Q_{x_1}').$$

	\noindent Next, let $P_{\theta}(x_1,\alpha)$ be the $8\times 1$ vector with elements $$\Pr(Y_{i2}=y_2,X_{i2}=x_2,Y_{i1}=y_1\,|\, X_{i1}=x_1,\alpha_i=\alpha),$$ for $(y_2,x_2,y_1)\in\{0,1\}^3$. Since $\widetilde{\phi}_{x_1}\in {\cal{N}}(\nabla_{\pi_{x_1}} Q_{x_1}')$, we have, for all $ \alpha\in {\cal{S}}$,
	$$ \widetilde{\phi}_{x_1}'P_{\theta}(x_1,\alpha)=\widetilde{\phi}_{x_1}'P_{\theta}(x_1,\underline{\alpha}_K)\equiv C_{x_1},$$
	where we have used the fact that $\pi_{x_1}(\underline{\alpha}_K)=1-\sum \limits_{k=1}^{K-1}\pi_{x_1}(\underline{\alpha}_k)$. 
	
	\noindent Let us define the following demeaned version of $\widetilde{\phi}_{x_1}$:\footnote{The $8\times 1$ vector $\phi_{x_1}$ represents a function $\phi_{x_1}:\{0,1\}^3\mapsto \mathbb{R}$. With some abuse of terminology we sometimes refer to $\phi_{x_1}$ as a vector and sometimes as a function.}
	$$\phi_{x_1}=\widetilde{\phi}_{x_1}-C_{x_1}\iota .$$

	\noindent Note that, since $\widetilde{\phi}_{x_1}$ is not constant, it follows that $\phi_{x_1}\neq 0$. Moreover, using (\ref{eq_notinrange_x1}) and (\ref{eq_Q_sumto1}) we have
	$${\phi}_{x_1} \in{\cal{N}}(\nabla_{\pi_{x_1}} Q_{x_1}')\cap {\cal{N}}(\nabla_{G_{x_1}} Q_{x_1}'),$$
	from which it follows that
	$$\mathrm{(i)}\quad \nabla_{\pi_{x_1}} Q_{x_1}'\phi_{x_1}=0,\quad \mathrm{(ii)}\quad \nabla_{G_{x_1}} Q_{x_1}'\phi_{x_1}=0.$$

	\noindent We are now going to use (i) and (ii) to show (\ref{eq_1})-(\ref{eq_2}). From (ii) we get, for all $\alpha\in {\cal{S}}$,
	\begin{align*}&\pi_{x_1}(\alpha)\bigg(\phi_{x_1}(1,1,1)F(\theta +\alpha) F(\theta x_1+\alpha)-\phi_{x_1}(1,1,0)F(\alpha) F(\theta x_1+\alpha)\\
		& +\phi_{x_1}(1,0,1)[1-F(\theta +\alpha)] F(\theta x_1+\alpha)-\phi_{x_1}(1,0,0)[1-F(\alpha)] F(\theta x_1+\alpha)\bigg)=0,\\
		&\pi_{x_1}(\alpha)\bigg(\phi_{x_1}(0,1,1)F(\theta +\alpha) [1-F(\theta x_1+\alpha)]-\phi_{x_1}(0,1,0)F(\alpha) [1-F(\theta x_1+\alpha)]\\
		& +\phi_{x_1}(0,0,1)[1-F(\theta +\alpha)] [1-F(\theta x_1+\alpha)]-\phi_{x_1}(0,0,0)[1-F(\alpha)] [1-F(\theta x_1+\alpha)]\bigg)=0.
	\end{align*}
	
	\noindent This implies, using Assumption \ref{ass_1},
	\begin{align*}&\phi_{x_1}(1,1,1)F(\theta +\alpha) -\phi_{x_1}(1,1,0)F(\alpha)  +\phi_{x_1}(1,0,1)[1-F(\theta +\alpha)]-\phi_{x_1}(1,0,0)[1-F(\alpha)]=0,\\
		&\phi_{x_1}(0,1,1)F(\theta +\alpha) -\phi_{x_1}(0,1,0)F(\alpha) +\phi_{x_1}(0,0,1)[1-F(\theta +\alpha)] -\phi_{x_1}(0,0,0)[1-F(\alpha)] =0,
	\end{align*}
	which coincides with (\ref{eq_1}).
	
	\noindent Lastly, from (i) we get, for all $ \alpha\in {\cal{S}}$,
	\begin{align*}
		\phi_{x_1}'P_{\theta}(x_1,\alpha)&=\phi_{x_1}'P_{\theta}(x_1,\underline{\alpha}_K) \\
		&=\widetilde{\phi}_{x_1}'P_{\theta}(x_1,\underline{\alpha}_K)-C_{x_1}\underbrace{\iota'P_{\theta}(x_1,\underline{\alpha}_K)}_{=1} \\
		&=\widetilde{\phi}_{x_1}'P_{\theta}(x_1,\underline{\alpha}_K)-\widetilde{\phi}_{x_1}'P_{\theta}(x_1,\underline{\alpha}_K) \\
		&=0,
	\end{align*}
	which can be equivalently written as
	$$\sum_{y_2=0}^1\sum_{x_2=0}^1\sum_{y_1=0}^1\phi_{x_1}(y_1,y_2,x_2)\Pr(Y_{i2}=y_2,X_{i2}=x_2,Y_{i1}=y_1\,|\, X_{i1}=x_1,\alpha_i=\alpha;\theta)=0.$$
	Now, using (\ref{eq_1}), this implies that, for all $x_2\in\{0,1\}$,
	$$\sum_{y_2=0}^1\sum_{y_1=0}^1\phi_{x_1}(y_1,y_2,x_2)\Pr(Y_{i2}=y_2\,|\, X_{i2}=x_2,\alpha_i=\alpha;\theta)\Pr(Y_{i1}=y_1\,|\, X_{i1}=x_1,\alpha_i=\alpha;\theta)=0,$$
	which coincides with (\ref{eq_2}).

	\section{Proof of Corollary \ref{coro_1}}

	The proof is by contradiction. Suppose that $\theta$ is point-identified. Then by (\ref{eq_1}) we have, for some $x_1\in\{0,1\}$, and for all $y_1\in\{0,1\}$ and $\alpha\in {\cal{S}}$,
	\begin{align*}&\phi_{x_1}(y_1,0,1)[1-F(\theta +\alpha)]+\phi_{x_1}(y_1,1,1)F(\theta +\alpha)=\phi_{x_1}(y_1,0,0)[1-F(\alpha)]+\phi_{x_1}(y_1,1,0)F(\alpha).
	\end{align*}
	Since $1$, $F(\alpha)$, and $F(\theta+\alpha)$, for $\alpha\in {\cal{S}}$, are linearly independent, we thus have, for all $y_1\in\{0,1\}$,
	\begin{equation}\phi_{x_1}(y_1,0,1)=\phi_{x_1}(y_1,1,1)=\phi_{x_1}(y_1,0,0)=\phi_{x_1}(y_1,1,0).\label{eq_equali}\end{equation}
	
	\noindent Next, using (\ref{eq_2}) at $x_2=1$ we have
	\begin{align*}
		&\phi_{x_1}(1,1,1)F(\theta+\alpha)F(\theta x_1+\alpha)+\phi_{x_1}(0,1,1)F(\theta+\alpha)[1-F(\theta x_1+\alpha)]\\
		&+\phi_{x_1}(1,0,1)[1-F(\theta+\alpha)]F(\theta x_1+\alpha)+\phi_{x_1}(0,0,1)[1-F(\theta+\alpha)][1-F(\theta x_1+\alpha)]=0.
	\end{align*}
	Using (\ref{eq_equali}) then gives
	\begin{align*}
		&\phi_{x_1}(1,1,1)F(\theta x_1+\alpha)+\phi_{x_1}(0,1,1)[1-F(\theta x_1+\alpha)]=0.
	\end{align*}
	Now, since $1$ and $F(\theta x_1+\alpha)$, for $\alpha\in {\cal{S}}$, are linearly independent, it follows that $$\phi_{x_1}(1,1,1)=\phi_{x_1}(0,1,1)=0.$$ Using (\ref{eq_equali}) then also gives $$\phi_{x_1}(1,0,1)=\phi_{x_1}(0,0,1)=0.$$ 
	
	\noindent Lastly, repeating the same argument starting with (\ref{eq_2}) at $x_2=0$ gives
	$$\phi_{x_1}(1,1,0)=\phi_{x_1}(0,1,0)=\phi_{x_1}(1,0,0)=\phi_{x_1}(0,0,0)=0.$$
	
	\noindent It follows that $\phi_{x_1}=0$, which leads to a contradiction. 
	
	\section{Proof of remark \ref{remark: sign_identification} (sign identification of $\theta$)} \label{appendix_section_sgntheta}
	Note that
	\begin{align} \label{identifying_moment_sgntheta}
		\begin{split}
			\mathbb{E}\left[Y_{i2}-Y_{i1}\,|\,X_{i1}=0\right]&= \mathbb{E}\left[\mathbb{E}\left[Y_{i2}\,|\,X_{i2},Y_{i1},X_{i1}=0,\alpha_i\right]-\mathbb{E}\left[Y_{i1}\,|\,X_{i1}=0,\alpha_i\right]\,|\,X_{i1}=0\right] \\ 
			&=\mathbb{E}\left[F(\theta X_{i2}+\alpha_i)-F(\alpha_i)\,|\,X_{i1}=0\right] \\
			&=\mathbb{E}\left[(F(\theta+\alpha_i)-F(\alpha_i))X_{i2}Y_{i1}+(F(\theta+\alpha_i)-F(\alpha_i))X_{i2}(1-Y_{i1})\,|\,X_{i1}=0\right] \\
			&=\int_{\mathcal{S}} \sum_{y_1=0}^1 (F(\theta+\alpha)-F(\alpha))\underbrace{G^2_{y_1,0}(\alpha)F(\alpha)^{y_1}(1-F(\alpha))^{1-y_1}\pi_{0}(\alpha)}_{>0 \text{ by Assumption } \ref{ass_1}}d\mu(\alpha).
		\end{split}
	\end{align}
	If $\theta=0$, (\ref{identifying_moment_sgntheta}) implies that $\mathbb{E}\left[Y_{i2}-Y_{i1}\,|\,X_{i1}=0\right]=0$. Moreover, since $F(\cdot)$ is strictly increasing, it follows that $\theta>0$ (respectively, $<0$) and $\mathbb{E}\left[Y_{i2}-Y_{i1}\,|\,X_{i1}=0\right]>0$ (resp., $<0$) are equivalent. This implies that $\mbox{sign}(\theta)=\mbox{sign}\left(\mathbb{E}\left[Y_{i2}-Y_{i1}\,|\,X_{i1}=0\right]\right)$. A similar argument applied to $X_{i1}=1$ implies that $\mbox{sign}(\theta)=\mbox{sign}\left(\mathbb{E}\left[Y_{i1}-Y_{i2}\,|\,X_{i1}=1\right]\right)$.
	
	\section{Identification in the exponential model\label{App_expo}}
	Let $$\overline{\phi}_{x_1}(\widetilde{\theta})\overset{def}{\equiv}\mathbb{E}[\phi_{x_1}(Y_1,Y_2,X_2;\widetilde{\theta})\,|\, X_{i1}=x_1]=\mathbb{E}[(1-Y_{i2})e^{\widetilde{\theta} X_{i2}}-(1-Y_{i1})e^{\widetilde{\theta} X_{i1}}\,|\, X_{i1}=x_1].$$ We show that $\theta$ is the unique solution to the equation 
	\begin{align*}
		\overline{\phi}_{x_1}(\widetilde{\theta})=0.
	\end{align*}
	Since $\overline{\phi}_{x_1}(\theta)=0$, the result will follow if one can show that, for any $x_{1}\in \{0,1\}$, $ \overline{\phi}_{x_1}$ is strictly monotonic. \\
	
	\noindent Let $(\theta_1,\theta_2)\in \Theta^2$ with $\theta_1>\theta_2$. For $x_1=0$, we have
	\begin{align*}
		\overline{\phi}_{0}(\theta_1)-\overline{\phi}_{0}(\theta_2)&=\mathbb{E}[(1-Y_{i2})e^{\theta_1 X_{i2}}-(1-Y_{i1})\,|\, X_{i1}=0] -\mathbb{E}[(1-Y_{i2})e^{\theta_2 X_{i2}}-(1-Y_{i1})\,|\, X_{i1}=0]\\
		&=\mathbb{E}[(1-Y_{i2})(e^{\theta_1 X_{i2}}-e^{\theta_2 X_{i2}})\,|\, X_{i1}=0] \\
		&=(e^{\theta_1}-e^{\theta_2})\mathbb{E}[(1-Y_{i2})X_{i2}\,|\, X_{i1}=0] \\
		&=(e^{\theta_1}-e^{\theta_2})\mathbb{E}[(1-F(\theta +\alpha_i))X_{i2}\,|\, X_{i1}=0] \\
		&=\underbrace{(e^{\theta_1}-e^{\theta_2})}_{>0} \int_{{\cal{S}}} \sum_{y_1=0}^1 \underbrace{(1-F(\theta +\alpha))G^{2}_{y_1,0}(\alpha)F(\alpha)^{y_1}(1-F(\alpha))^{1-y_{1}}\pi_{0}(\alpha)}_{>0 \text{ by Assumption \ref{ass_1}}}d\mu(\alpha)
		\\
		&>0,
	\end{align*}
	which shows that $\overline{\phi}_{0}$ is strictly increasing. If $x_1=1$, then
	\begin{align*}
		\overline{\phi}_{1}(\theta_1)-\overline{\phi}_{1}(\theta_2)&=\mathbb{E}[(1-Y_{i2})e^{\theta_1 X_{i2}}-(1-Y_{i1})e^{\theta_1}\,|\, X_{i1}=1] -\mathbb{E}[(1-Y_{i2})e^{\theta_2 X_{i2}}-(1-Y_{i1})e^{\theta_2}\,|\, X_{i1}=1]\\
		&=\mathbb{E}[(1-Y_{i2})(e^{\theta_1 X_{i2}}-e^{\theta_2 X_{i2}})-(1-Y_{i1})(e^{\theta_1}-e^{\theta_2})\,|\, X_{i1}=1] \\
		&=(e^{\theta_1}-e^{\theta_2})\mathbb{E}[(1-Y_{i2})X_{i2}-(1-Y_{i1})\,|\, X_{i1}=1] \\
		&=-(e^{\theta_1}-e^{\theta_2})\mathbb{E}[(1-F(\theta +\alpha_i))(1-X_{i2})\,|\, X_{i1}=1] \\
		&=-\underbrace{(e^{\theta_1}-e^{\theta_2})}_{>0}\times \\
		&\int_{{\cal{S}}} \sum_{y_1=0}^1 \underbrace{(1-F(\theta+\alpha))(1-G^{2}_{y_1,1}(\alpha))F(\theta+\alpha)^{y_1}(1-F(\theta+\alpha))^{1-y_{1}}\pi_{1}(\alpha)}_{>0 \text{ by Assumption \ref{ass_1}}}d\mu(\alpha) \\
		&<0,
	\end{align*}
	which shows that $\overline{\phi}_{1}$ is strictly decreasing. 
	
	\section{Proof of Lemma \ref{lem_2}}
	
	In what follows we assume $T\geq 3$, having already proved the validity of the claim for $T=2$ in Lemma \ref{lem_1}. \\
	
	\noindent Since $\theta$ is point-identified it is locally point-identified. Additionally, since $(\theta,\pi,G)$ is a regular point of $\nabla Q(\theta,\pi,G)$ by Assumption \ref{ass_2}, we can appeal to Theorem 8 in \cite{bekker2001identification} and follow the same line of arguments as in the proof of Lemma \ref{lem_1} to conclude that 
	there exists $x_1\in\{0,1\}$ and a $2^{2T-1}\times 1$ vector $\phi_{x_1}\neq 0$ such that
	$$\mathrm{(i)}\quad \nabla_{\pi_{x_1}} Q_{x_1}'\phi_{x_1}=0,\quad \mathrm{(ii)}\quad \nabla_{G_{x_1}} Q_{x_1}'\phi_{x_1}=0.$$

	\noindent We will now prove (\ref{eq_1_general}) and (\ref{eq_2_general}) using  finite induction. 
	
	\noindent Let us start with (\ref{eq_1_general}). Given $s\in\{0,...,T-2\}$, let $\mathcal{P}(s)$ denote the statement that,
	for all $y^{T-(s+1)}\in \{0,1\}^{T-(s+1)}$ and $x^{T-(s+1)}\in \{0,1\}^{T-(s+1)}$,
	\begin{align*}
		\sum_{y^{T-s:T}\in \{0,1\}^{s+1}} \phi_{x_1}(y^{T},x^{2:T})\prod_{t=T-s}^T F(\theta x_{t}+\alpha)^{y_t}[1-F(\theta x_{t}+\alpha)]^{1-y_t}
	\end{align*}
	does not depend on  $x^{T-s:T}$. 
	\\
	\newline
	\textbf{Base case:}
	
	\noindent Condition (ii) implies that
	\begin{align*}
		\left(	\frac{\partial Q_{x_1}}{\partial G^T_{y^{T-1},x^{T-1}}(\alpha)}\right)'\phi_{x_1}=0,
	\end{align*}
	or equivalently that
	\begin{align*}
		&\sum_{y_T=0}^1 \sum_{x_T=0}^1 \phi_{x_1}(y^{T},x^{2:T}) F(\theta x_T+\alpha)^{y_T}[1-F(\theta x_T+\alpha)]^{1-y_T} (-1)^{1-x_T}  \\
		&\times\prod_{t=2}^{T-1} F(\theta x_t+\alpha)^{y_t}[1-F(\theta x_t+\alpha)]^{1-y_t} G^t_{y^{t-1},x^{t-1}}(\alpha)^{x_t}[1-G^t_{y^{t-1},x^{t-1}}(\alpha)]^{1-x_t} \\
		& \times F(\theta x_1+\alpha)^{y_1}[1-F(\theta x_1+\alpha)]^{1-y_1} =0 .
	\end{align*}
	Using Assumption \ref{ass_1}, this simplifies to
	\begin{align*}
		\sum_{y_T=0}^1 \sum_{x_T=0}^1 \phi_{x_1}(y^{T},x^{2:T}) F(\theta x_T+\alpha)^{y_T}[1-F(\theta x_T+\alpha)]^{1-y_T} (-1)^{1-x_T}=0,
	\end{align*}
	which implies that
	\begin{align*}
		&	\sum_{y_T=0}^1 \phi_{x_1}(y^{T},x^{2:T}) F(\theta x_T+\alpha)^{y_T}[1-F(\theta x_T+\alpha)]^{1-y_T} \end{align*}
	does not depend on $x_T$. 
	
	\noindent Thus, $\mathcal{P}(0)$ is true.
	\\
	\newline
	\textbf{Induction step:} \\
	Suppose that $\mathcal{P}(0),\ldots,\mathcal{P}(s)$ are true for $s\in\{0,\ldots,T-3\}$. We are going to show that $\mathcal{P}(s+1)$ is true. \\
	\noindent Condition (ii) implies that
	\begin{align*}
		\left(\frac{\partial Q_{x_1}}{\partial G^{T-(s+1)}_{y^{T-(s+2)},x^{T-(s+2)}}(\alpha)}\right)'\phi_{x_1}=0.
	\end{align*}
	If $s<(T-3)$, this corresponds to
	\begin{align*}
		&\sum_{y^{T-(s+1):T}\in \{0,1\}^{s+2}} \sum_{x^{T-(s+1):T}\in \{0,1\}^{s+2}}\phi_{x_1}(y^{T},x^{2:T})\\ 
		& \times \prod_{t=T-s}^{T}F(\theta x_t+\alpha)^{y_t}[1-F(\theta x_t+\alpha)]^{1-y_t}G^t_{y^{t},x^{t}}(\alpha)^{x_t}[1-G^t_{y^{t},x^{t}}(\alpha)]^{1-x_t}  \\
		& \times F(\theta x_{T-(s+1)}+\alpha)^{y_{T-(s+1)}}[1-F(\theta x_{T-(s+1)}+\alpha)]^{1-y_{T-(s+1)}} (-1)^{1-x_{T-(s+1)}}  \\
		&\times\prod_{t=2}^{T-(s+2)} F(\theta x_t+\alpha)^{y_t}[1-F(\theta x_t+\alpha)]^{1-y_t} G^t_{y^{t-1},x^{t-1}}(\alpha)^{x_t}[1-G^t_{y^{t-1},x^{t-1}}(\alpha)]^{1-x_t}  \\
		&\times F(\theta x_1+\alpha)^{y_1}[1-F(\theta x_1+\alpha)]^{1-y_1} =0. 
	\end{align*}
	While if $s=(T-3)$, this corresponds to
	\begin{align*}
		&\sum_{y^{2:T}\in \{0,1\}^{T-1}} \sum_{x^{2:T}\in \{0,1\}^{T-1}}\phi_{x_1}(y^{T},x^{2:T})  \\ 
		&\times\prod_{t=3}^{T}F(\theta x_t+\alpha)^{y_t}[1-F(\theta x_t+\alpha)]^{1-y_t}G^t_{y^{t},x^{t}}(\alpha)^{x_t}[1-G^t_{y^{t},x^{t}}(\alpha)]^{1-x_t} \\
		&\times F(\theta x_{2}+\alpha)^{y_{2}}[1-F(\theta x_{2}+\alpha)]^{1-y_{2}} (-1)^{1-x_{2}} \\
		&\times F(\theta x_1+\alpha)^{y_1}[1-F(\theta x_1+\alpha)]^{1-y_1} =0. 
	\end{align*}
	Using Assumption \ref{ass_1} this gives, for all $s\in\{0,\ldots,T-3\}$,
	\begin{align}
		&\sum_{y^{T-(s+1):T}\in \{0,1\}^{s+2}} \sum_{x^{T-(s+1):T}\in \{0,1\}^{s+2}}\phi_{x_1}(y^{T},x^{2:T})\notag  \\ 
		&\times\prod_{t=T-s}^{T}F(\theta x_t+\alpha)^{y_t}[1-F(\theta x_t+\alpha)]^{1-y_t}G^t_{y^{t},x^{t}}(\alpha)^{x_t}[1-G^t_{y^{t},x^{t}}(\alpha)]^{1-x_t}  \notag\\
		&\times F(\theta x_{T-(s+1)}+\alpha)^{y_{T-(s+1)}}[1-F(\theta x_{T-(s+1)}+\alpha)]^{1-y_{T-(s+1)}} (-1)^{1-x_{T-(s+1)}} =0 .\label{eq_splus1}
	\end{align}
	
	\noindent Let $L_{s+1}$ denote the left-hand side of (\ref{eq_splus1}). Exploiting successively the fact that $\mathcal{P}(0),\ldots,\mathcal{P}(s)$ are true, alongside the property that, for all $ t\in\{T-s,...,T\}$,
	\begin{align}
		\sum_{x_t=0}^1G^t_{y^{t},x^{t}}(\alpha)^{x_t}[1-G^t_{y^{t},x^{t}}(\alpha)]^{1-x_t}=1,\label{eq_sumG}
	\end{align}
	it is easy to see that 
	\begin{align*}
		L_{s+1}&=\sum_{y^{T-(s+1):T}\in \{0,1\}^{s+2}} \sum_{x_{T-(s+1)}=0}^1\phi_{x_1}(y^{T},x^{2:T}) \prod_{t=T-s}^{T}F(\theta x_t+\alpha)^{y_t}[1-F(\theta x_t+\alpha)]^{1-y_t}  \\
		&\times F(\theta x_{T-(s+1)}+\alpha)^{y_{T-(s+1)}}[1-F(\theta x_{T-(s+1)}+\alpha)]^{1-y_{T-(s+1)}} (-1)^{1-x_{T-(s+1)}} =0.
	\end{align*}
	
	\noindent Recalling that $\mathcal{P}(s)$ is true, this implies that 
	\begin{align*}
		\sum_{y^{T-(s+1):T}\in \{0,1\}^{s+1}} \phi_{x_1}(y^{T},x^{2:T})\prod_{t=T-(s+1)}^T F(\theta x_{t}+\alpha)^{y_t}[1-F(\theta x_{t}+\alpha)]^{1-y_t}
	\end{align*}
	does not depend on $x^{T-(s+1):T}$. Hence, $\mathcal{P}(s+1)$ is true. This concludes the proof of (\ref{eq_1_general}).  \\

	\noindent 	Finally, we show (\ref{eq_2_general}). As in the proof of Lemma \ref{lem_1}, Condition (i) implies that
	\begin{align*}
		&\sum_{y^{T}\in \{0,1\}^{T}}\sum_{x^{2:T}\in \{0,1\}^{T-1}}\phi_{x_1}(y^{T},x^{2:T}) \\
		&\times \prod_{t=2}^{T}F(\theta x_t+\alpha)^{y_t}[1-F(\theta x_t+\alpha)]^{1-y_t}G^t_{y^{t},x^{t}}(\alpha)^{x_t}[1-G^t_{y^{t},x^{t}}(\alpha)]^{1-x_t}\\
		&\times F(\theta x_1+\alpha)^{y_1}[1-F(\theta x_1+\alpha)]^{1-y_1}
		=0.\end{align*}
	Using (\ref{eq_1_general}) and (\ref{eq_sumG}), it follows that
	\begin{align*}
		&\sum_{y^{T}\in \{0,1\}^{T}}\phi_{x_1}(y^{T},x^{2:T}) \prod_{t=1}^{T}F(\theta x_t+\alpha)^{y_t}[1-F(\theta x_t+\alpha)]^{1-y_t}
		=0,\end{align*}
	which coincides with (\ref{eq_2_general}).

	\section{Proof of Corollary \ref{coro_3}}
	
	In what follows we assume $T\geq 3$, having already proved the validity of the claim for $T=2$ in Corollary \ref{coro_1}. \\
	
	\noindent The proof is by contradiction. Suppose that $\theta$ is point-identified. We will show that this necessarily leads to $\phi_{x_1}=0$, which will contradict Lemma \ref{lem_2}. To that end, we will first prove via finite induction that $\phi_{x_1}$ must be a constant function. \\
	\newline 
	\noindent For $s\in\{1,...,T-2\}$, let $\mathcal{P}(s)$ denote the statement that there exists a function $\phi_{x_1}^{T-s}:\{0,1\}^{2T-2s-1}\rightarrow \mathbb{R}$ such that, for all $y^{T}\in \{0,1\}^{T}$ and $x^{2:T}\in \{0,1\}^{T-1}$, we have $$\phi_{x_1}(y^{T},x^{2:T})=\phi_{x_1}^{T-s}(y^{T-s},x^{2:T-s}).$$
	\\
	\noindent \textbf{Base case:}
	\\
	\noindent By (\ref{eq_1_general}), the quantity
	\begin{align}
		\sum_{y_{T}=0}^1 \phi_{x_1}(y^{T},x^{2:T}) F(\theta x_{T}+\alpha)^{y_T}[1-F(\theta x_{T}+\alpha)]^{1-y_T} 
	\end{align}
	does not depend on $x_{T}$. Hence
	\begin{align*}&\phi_{x_1}(y^{T-1},1,x^{2:T-1},1)F(\theta+\alpha)+\phi_{x_1}(y^{T-1},0,x^{2:T-1},1)[1-F(\theta+\alpha)]\\
		&=\phi_{x_1}(y^{T-1},1,x^{2:T-1},0)F(\alpha)+\phi_{x_1}(y^{T-1},0,x^{2:T-1},0)[1-F(\alpha)].
	\end{align*}
	By linear independence of $1$, $F(\alpha)$, and $F(\theta+\alpha)$, this implies that 
	$\phi_{x_1}(y^{T},x^{2:T})$ does not depend on $(y_T,x_T)$. Hence $\mathcal{P}(1)$ is true.
	\\
	\newline
	\noindent \textbf{Induction step}
	\\
	\newline
	\noindent Suppose that $\mathcal{P}(s)$ is true for $s\in\{1,...,T-3\}$. Let us show that $\mathcal{P}(s+1)$ is true. \\
	\newline \noindent Since $\mathcal{P}(s)$ is true, we know that 
	there exists  a function $\phi_{x_1}^{T-s}:\{0,1\}^{2T-2s-1}\rightarrow \mathbb{R}$ such that $$\phi_{x_1}(y^{T},x^{2:T})=\phi_{x_1}^{T-s}(y^{T-s},x^{2:T-s}).$$
	
	\noindent Thus, by (\ref{eq_1_general}), the quantity:
	\begin{align*}
		& \sum_{y^{T-s:T}\in \{0,1\}^{s+1}} \phi_{x_1}(y^{T},x^{2:T})\prod_{t=T-s}^T F(\theta x_{t}+\alpha)^{y_t}[1-F(\theta x_{t}+\alpha)]^{1-y_t} \\
		&= \sum_{y_{T-s}=0}^1 \phi_{x_1}^{T-s}(y^{T-s},x^{2:T-s})\sum_{y^{T-(s-1):T}\in \{0,1\}^{s}} \prod_{t=T-(s-1)}^T F(\theta x_{t}+\alpha)^{y_t}[1-F(\theta x_{t}+\alpha)]^{1-y_t}  \\
		&\times F(\theta x_{T-s}+\alpha)^{y_{T-s}}[1-F(\theta x_{T-s}+\alpha)]^{1-y_{T-s}} \\
		&=\sum_{y_{T-s}=0}^1 \phi_{x_1}^{T-s}(y^{T-s},x^{2:T-s})F(\theta x_{T-s}+\alpha)^{y_{T-s}}[1-F(\theta x_{T-s}+\alpha)]^{1-y_{T-s}}
	\end{align*}
	does not depend on $x^{T-s:T}$. Therefore, 
	\begin{align*}
		&\phi_{x_1}^{T-s}(y^{T-s-1},1,x^{2:T-s-1},1)F(\theta+\alpha)+\phi_{x_1}^{T-s}(y^{T-s-1},0,x^{2:T-s-1},1)[1-F(\theta+\alpha)]\\
		&=\phi_{x_1}^{T-s}(y^{T-s-1},1,x^{2:T-s-1},0)F(\alpha)+\phi_{x_1}^{T-s}(y^{T-s-1},0,x^{2:T-s-1},0)[1-F(\alpha)].
	\end{align*}
	Since $1$, $F(\alpha)$, and $F(\theta+\alpha)$ are linearly independent, this implies $\mathcal{P}(s+1)$. \\
	It follows from the previous induction argument that there exists a function $\phi_{x_1}^{2}:\{0,1\}^{3}\rightarrow \mathbb{R}$ such that, for all $(y^{T},x^{2:T})$, 	\begin{align*}
		\phi_{x_1}(y^{T},x^{2:T})=\phi_{x_1}^{2}(y^{2},x_{2}).
	\end{align*}
	Using (\ref{eq_1_general}), the quantity
	\begin{align*}
		&\sum_{y^{2:T}\in \{0,1\}^{T-1}} \phi_{x_1}(y^{T},x^{2:T})\prod_{t=2}^T F(\theta x_{t}+\alpha)^{y_t}[1-F(\theta x_{t}+\alpha)]^{1-y_t} \\
		&=\sum_{y_{2}=0}^1 \phi_{x_1}^{2}(y^{2},x_{2})F(\theta x_{2}+\alpha)^{y_2}[1-F(\theta x_{2}+\alpha)]^{1-y_2}
	\end{align*}
	does not depend on $x^{2:T}$. Therefore, 
	\begin{align*}
		&\phi_{x_1}^{2}(y_1,1,1)F(\theta+\alpha)+\phi_{x_1}^{2}(y_1,0,1)[1-F(\theta+\alpha)]\\
		&=\phi_{x_1}^{2}(y_1,1,0)F(\alpha)+\phi_{x_1}^{2}(y_1,0,0)[1-F(\alpha)].
	\end{align*}
	Since $1$, $F(\alpha)$, and $F(\theta+\alpha)$ are linearly independent, this implies that there exists a function $\phi_{x_1}^{1}:\{0,1\}\rightarrow \mathbb{R}$ such that, for all $(y^T,x^{2:T})$, 
	\begin{align*}
		\phi_{x_1}(y^{T},x^{2:T})=\phi_{x_1}^{1}(y_{1}).
	\end{align*}
	Lastly, (\ref{eq_2_general}) implies
	\begin{align*}
		&\sum_{y^{T}\in \{0,1\}^{T}} \phi_{x_1}(y^{T},x^{2:T})\prod_{t=1}^T F(\theta x_{t}+\alpha)^{y_t}[1-F(\theta x_{t}+\alpha)]^{1-y_t} \\
		&=\sum_{y^{T}\in \{0,1\}^{T}} \phi_{x_1}^{1}(y_{1})\prod_{t=1}^T F(\theta x_{t}+\alpha)^{y_t}[1-F(\theta x_{t}+\alpha)]^{1-y_t} \\
		&=\sum_{y_1=0}^1 \phi_{x_1}^{1}(y_{1}) \sum_{y^{2:T}\in \{0,1\}^{T}} \prod_{t=1}^T F(\theta x_{t}+\alpha)^{y_t}[1-F(\theta x_{t}+\alpha)]^{1-y_t} \\
		&=\sum_{y_1=0}^1 \phi_{x_1}^{1}(y_{1})  F(\theta x_{1}+\alpha)^{y_1}[1-F(\theta x_{1}+\alpha)]^{1-y_1} \\
		&=0.
	\end{align*}
	Linear independence of $1$, $F(\alpha)$, and $F(\theta+\alpha)$ thus implies
	\begin{align*}
		\phi_{x_1}^{1}(0)=\phi_{x_1}^{1}(1)=0.
	\end{align*}
	Therefore, $\phi_{x_1}$ must be the null function, a contradiction.

	\section{Proof of Proposition \ref{lem_thetaI}\label{App_IS}}
	
	It is immediate to verify that, if $\widetilde{\theta}\in\Theta^I$, then (\ref{eq_IS1_T}), (\ref{eq_IS2_T}) and (\ref{eq_IS3_T}) are satisfied.
	
	\noindent Conversely, suppose that (\ref{eq_IS1_T}), (\ref{eq_IS2_T}) and (\ref{eq_IS3_T}) are satisfied. Let
	\begin{equation}p_{x_1}(y^{T},x^{2:T},\alpha)
		=F(\widetilde{\theta} x_T+\alpha)^{y_T}[1-F(\widetilde{\theta} x_T+\alpha)]^{1-y_T}\psi_{x_1}(x^{2:T},y^{T-1},\alpha).\label{eq_mu}\end{equation}

	\noindent Using (\ref{eq_IS2_T}) we have
	\begin{equation*}p_{x_1}(y^{T},x^{2:T},\alpha)\geq 0, \quad \sum_{y^{T}\in\{0,1\}^T}\sum_{x^{2:T}\in\{0,1\}^{T-1}}\int_{{\cal{S}}}p_{x_1}(y^{T},x^{2:T},\alpha)d\mu(\alpha)=1,\end{equation*}
	so $p_{x_1}$ is a valid distribution function (conditional on $X_{i1}=x_1$).
	
	\noindent Next, using (\ref{eq_IS1_T}) we have
	\begin{align*}&\int_{{\cal{S}}}p_{x_1}(y^{T},x^{2:T},\alpha)d\mu(\alpha)\\&=
		\int_{{\cal{S}}} F(\widetilde{\theta} x_T+\alpha)^{y_T}[1-F(\widetilde{\theta} x_T+\alpha)]^{1-y_T}\psi_{x_1}(x^{2:T},y^{T-1},\alpha)d\mu(\alpha)\\
		&=Q_{x_1}(y^{T},x^{2:T};\theta,\pi,G),\end{align*}
	so $p_{x_1}$ is consistent with the conditional distribution $Q_{x_1}(y^{T},x^{2:T};\theta,\pi,G)$ of $(Y_{i}^{T},X_i^{2:T})$ given $X_{i1}$.

	\noindent Next, using (\ref{eq_IS3_T}) we have, for all $s\in\{2,...,T\}$,
	\begin{align*}&\sum_{x^{s:T}\in\{0,1\}^{T-s+1}}\sum_{y^{s:T}\in\{0,1\}^{T-s+1}}p_{x_1}(y^{T},x^{2:T},\alpha)\\
		&=\sum_{x^{s:T}\in\{0,1\}^{T-s+1}}\sum_{y^{s:T-1}\in\{0,1\}^{T-s}}\left\{\sum_{y_{T}=0}^1 F(\widetilde{\theta} x_T+\alpha)^{y_T}[1-F(\widetilde{\theta} x_T+\alpha)]^{1-y_T}\right\}\psi_{x_1}(x^{2:T},y^{T-1},\alpha)\\
		&=\sum_{x^{s:T}\in\{0,1\}^{T-s+1}}\sum_{y^{s:T-1}\in\{0,1\}^{T-s}}\psi_{x_1}(x^{2:T},y^{T-1},\alpha)\\
		&= F(\widetilde{\theta} x_{s-1}+\alpha)^{y_{s-1}}[1-F(\widetilde{\theta} x_{s-1}+\alpha)]^{1-y_{s-1}}\sum_{x^{s:T}\in\{0,1\}^{T-s+1}}\sum_{y^{s-1:T-1}\in\{0,1\}^{T-s+1}}\psi_{x_1}(x^{2:T},y^{T-1},\alpha)\\
		&= F(\widetilde{\theta} x_{s-1}+\alpha)^{y_{s-1}}[1-F(\widetilde{\theta} x_{s-1}+\alpha)]^{1-y_{s-1}}\sum_{x^{s:T}\in\{0,1\}^{T-s+1}}\sum_{y^{s-1:T}\in\{0,1\}^{T-s+2}}p_{x_1}(x^{2:T},y^{T},\alpha),
	\end{align*}
	so, for all $t\in\{1,...,T-1\}$, the conditional distributions of $Y_{it}$ given $(Y_{i}^{t-1},X_i^{t-1},\alpha_i)$ induced by $p_{x_1}$ coincide with the ones under the model; i.e., with $F(\widetilde{\theta} x_{t}+\alpha)^{y_{t}}[1-F(\widetilde{\theta} x_{t}+\alpha)]^{1-y_{t}}$. 
	
	\noindent Lastly, using  (\ref{eq_mu}) we have
	\begin{align*}&p_{x_1}(y^{T},x^{2:T},\alpha)= F(\widetilde{\theta} x_T+\alpha)^{y_T}[1-F(\widetilde{\theta} x_T+\alpha)]^{1-y_T}\psi_{x_1}(x^{2:T},y^{T-1},\alpha)\\
		&=F(\widetilde{\theta} x_T+\alpha)^{y_T}[1-F(\widetilde{\theta} x_T+\alpha)]^{1-y_T}\sum_{y_T=0}^1p_{x_1}(y^{T},x^{2:T},\alpha),
	\end{align*}
	so the conditional distribution of $Y_{iT}$ given $(Y_{i}^{T-1},X_i^{T-1},\alpha_i)$ induced by $p_{x_1}$ also coincides with the one under the model.

	\noindent This implies that $\widetilde{\theta}\in\Theta^I$.
	
	\section{Computation of identified sets\label{App_implement}}
	In this section we describe the practical implementation of the linear programming approach for the computation of identified sets for two types of target parameters: $\theta$, and average partial effects. For simplicity of exposition we discuss the case  $T=2$, but the construction is analogous for larger $T$.
	
	\subsection{Parameter $\theta$}
	In Proposition \ref{lem_thetaI}, we established that a candidate parameter  $\widetilde{\theta}$ lies in the identified set $ \Theta^I$ if and only if one can find functions
	$\psi_0,\psi_1$ verifying equations (\ref{eq_IS2}), (\ref{eq_IS3}) and (\ref{eq_IS1}). A useful observation is that these conditions can be viewed as the constraints of a linear program. Thus, determining whether $\widetilde{\theta} \in \Theta^I$ is equivalent to determining the feasibility of a linear optimization problem. In the numerical illustration, we specifically consider:
	\begin{align*}
		\inf_{\psi_{0},\psi_{1}} \int_{{\cal{S}}} \sum_{x_1=0}^1 q_{x_1} \sum_{x_2=0}^1 \sum_{y_1=0}^1 \psi_{x_1}(x_2,y_1,\alpha) d\mu(\alpha),
	\end{align*}
	where the constraints are that $\psi_0,\psi_1$ satisfy equations (\ref{eq_IS2}), (\ref{eq_IS3}) and (\ref{eq_IS1}).  The additional constraints for the strictly exogenous case are that $\psi_0,\psi_1$ also verify the relationship presented in footnote 5.
	\subsection{Average partial effect $\Delta$}
	In addition to $\theta$, a quantity of interest is the average partial effect
	\begin{align*}
		\Delta&=\mathbb{E}[\Pr(Y_{i2}=1\,|\, X_{i2}=1,\alpha_i)-\Pr(Y_{i2}=1\,|\, X_{i2}=0,\alpha_i)]  \\
		&=\int_{{\cal{S}}}[F(\theta+\alpha)-F(\alpha)]\sum_{x_1\in\{0,1\}}q_{x_1}\pi_{x_1}(\alpha)d\mu(\alpha).
	\end{align*}
	which is generally not point-identified. Yet,  for a given $\widetilde{\theta}  \in \Theta^I$, one can compute a lower bound $\underline{\Delta}(\widetilde{\theta})$ and an upper bound $\overline{\Delta}(\widetilde{\theta})$ on the range of possible average partial effects as solutions to the following linear optimization problem:
	\begin{align*}
		\underline{\Delta}(\widetilde{\theta})&=\inf_{\psi_{0},\psi_{1}} \int_{{\cal{S}}}[F(\widetilde{\theta}+\alpha)-F(\alpha)]\sum_{x_1\in\{0,1\}}q_{x_1} \sum_{x_2\in\{0,1\}}\sum_{y_1\in\{0,1\}}\psi_{x_1}(x_2,y_1,\alpha) d\mu(\alpha),\\
		\overline{\Delta}(\widetilde{\theta})&=\sup_{\psi_{0},\psi_{1}} \int_{{\cal{S}}}[F(\widetilde{\theta}+\alpha)-F(\alpha)]\sum_{x_1\in\{0,1\}}q_{x_1} \sum_{x_2\in\{0,1\}}\sum_{y_1\in\{0,1\}} \psi_{x_1}(x_2,y_1,\alpha)d\mu(\alpha),
	\end{align*}
	subject to $\psi_0,\psi_1$ satisfying equations (\ref{eq_IS1}), (\ref{eq_IS2}), and (\ref{eq_IS3}). Under the assumption of strict exogeneity, $\psi_0$ and $\psi_1$ have to satisfy the additional constraint discussed in footnote 5. The sharp bounds for $\Delta$ are then obtained as
	\begin{align*}
		\underline{\Delta} &= \inf_{\widetilde{\theta}  \in \Theta^I}\underline{\Delta}(\widetilde{\theta}), \\
		\overline{\Delta} &= \sup_{\widetilde{\theta}  \in \Theta^I}\overline{\Delta}(\widetilde{\theta}).
	\end{align*}

	\clearpage	
	\begin{figure}[tbp]
		\caption{Approximate identified sets for logit and probit models with $T=2$\label{App_Fig_IS_rob}}
		\begin{center}
			\begin{tabular}{cc}
				LOGIT MODEL & PROBIT MODEL\\
				\multicolumn{2}{c}{$K=5$}   \\
				\includegraphics[width=80mm, height=55mm]{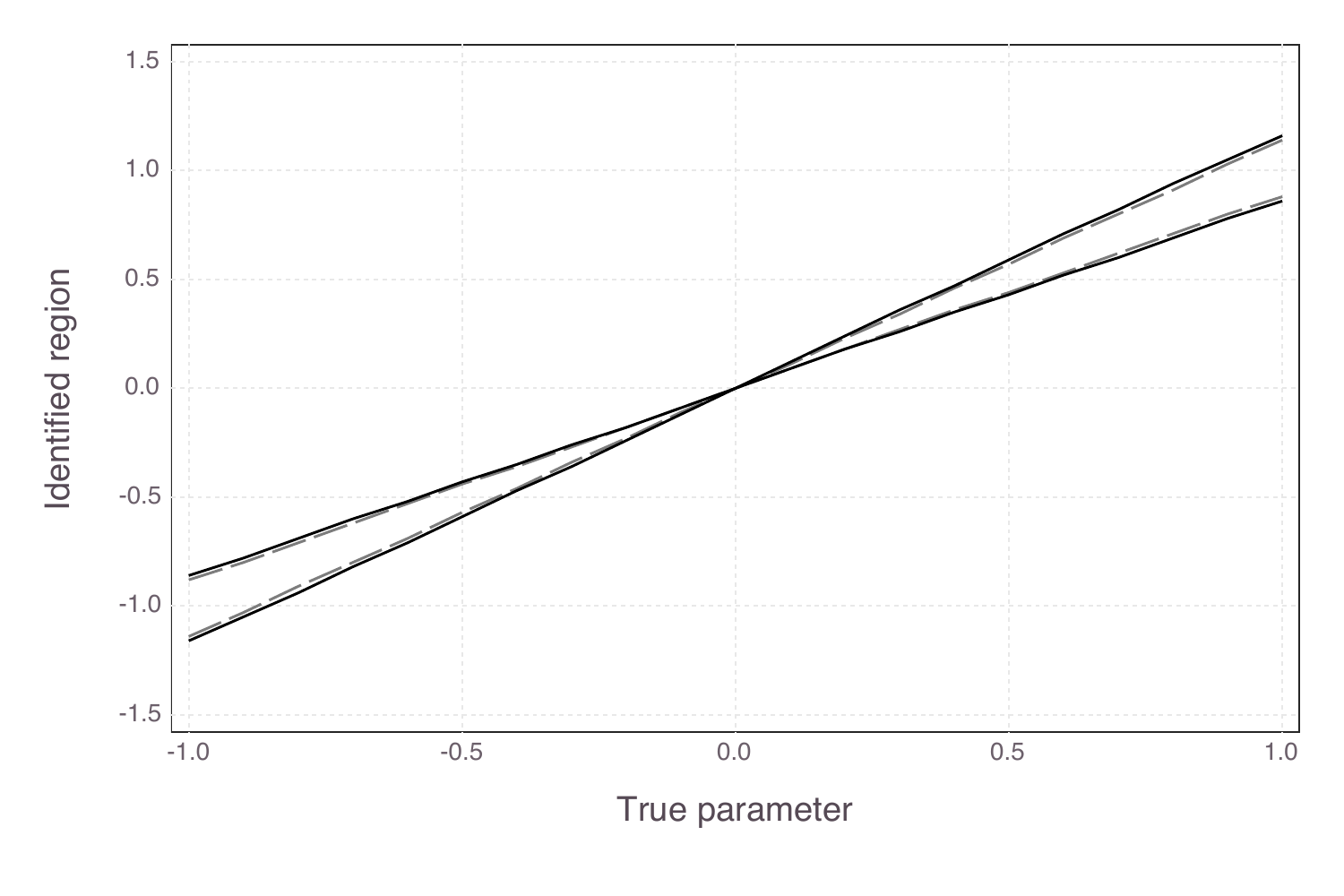} & \includegraphics[width=80mm, height=55mm]{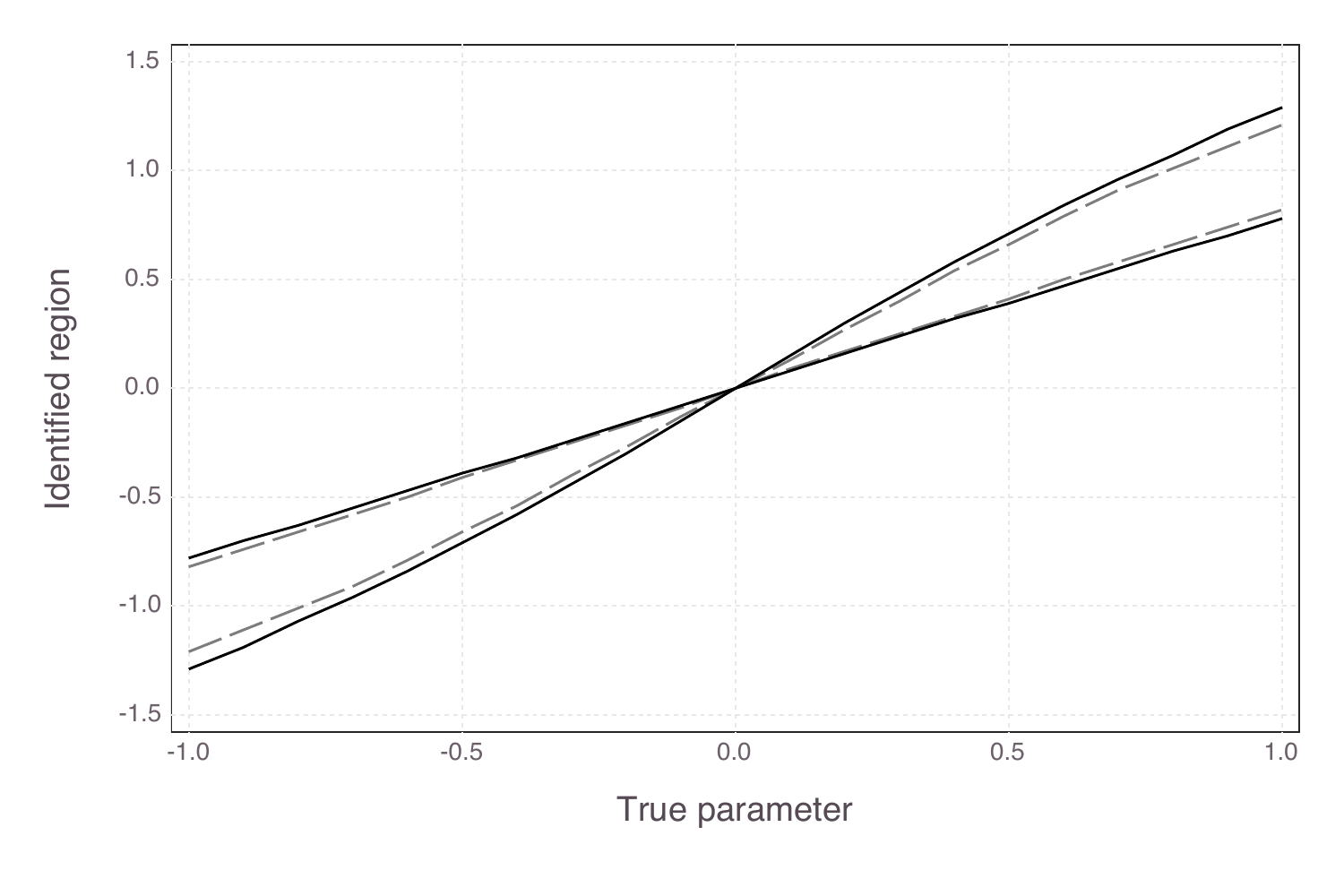} \\
				\multicolumn{2}{c}{$K=50$} \\  
				\includegraphics[width=80mm, height=55mm]{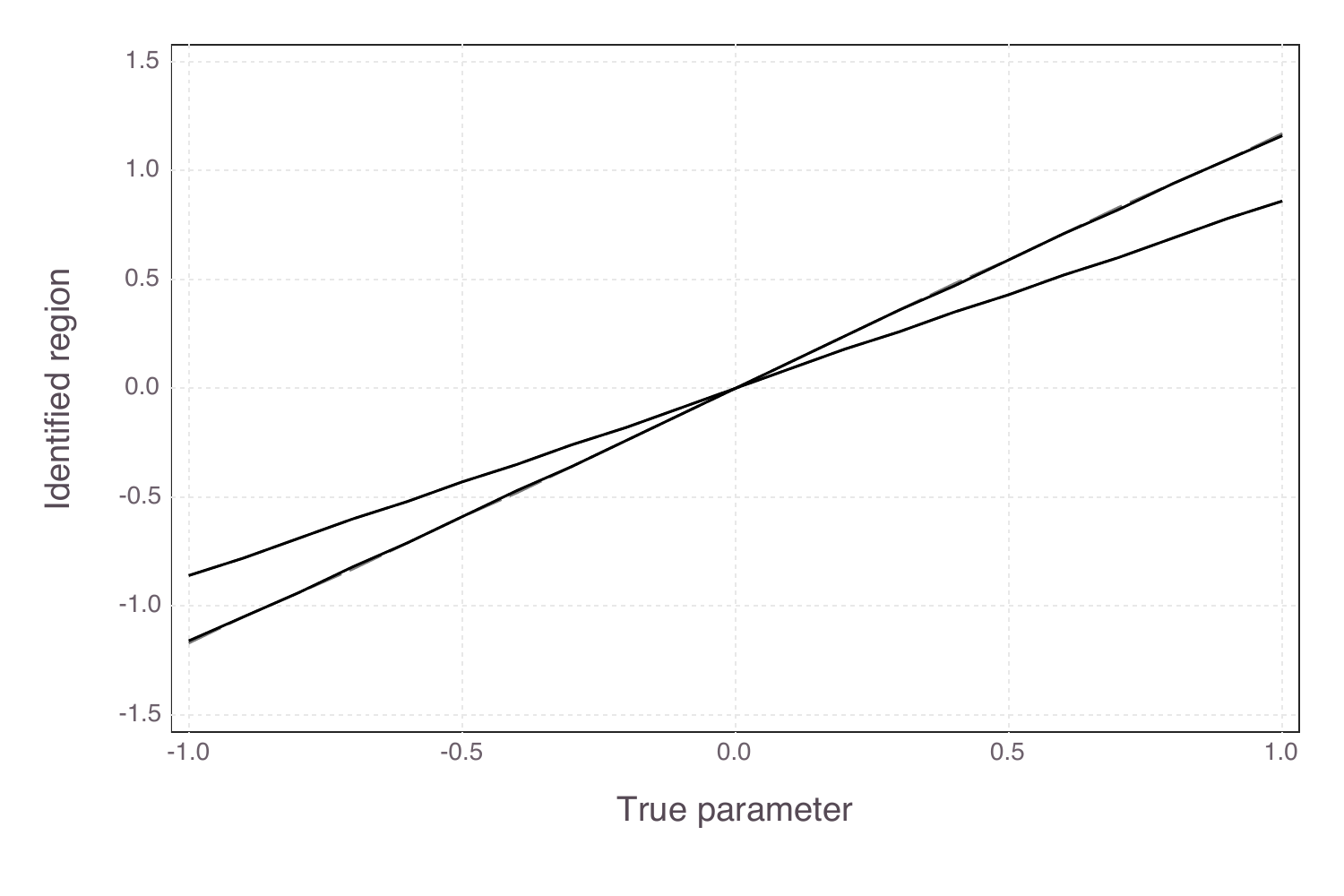} & \includegraphics[width=80mm, height=55mm]{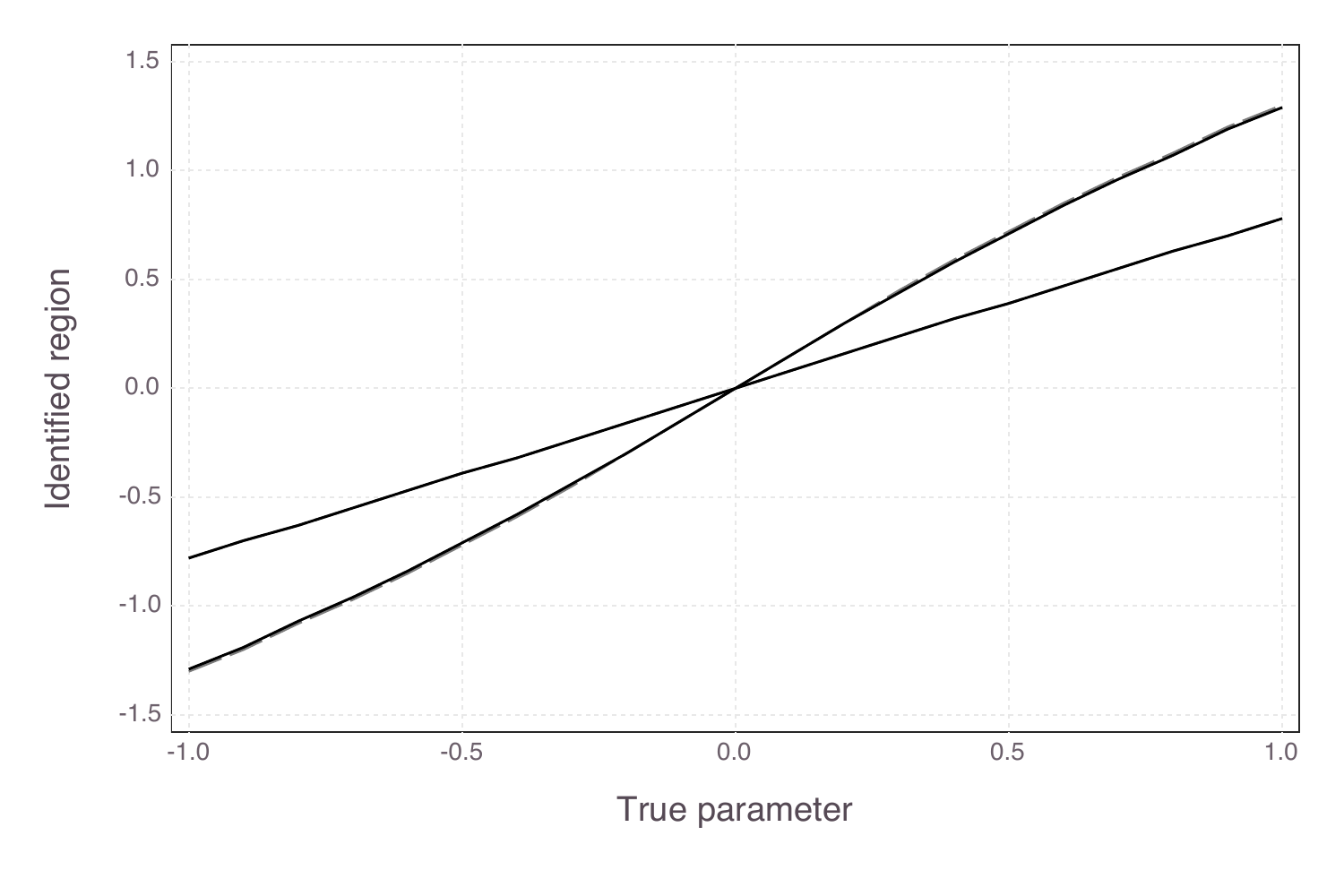} \\
				\multicolumn{2}{c}{$K=500$}\\  
				\includegraphics[width=80mm, height=55mm]{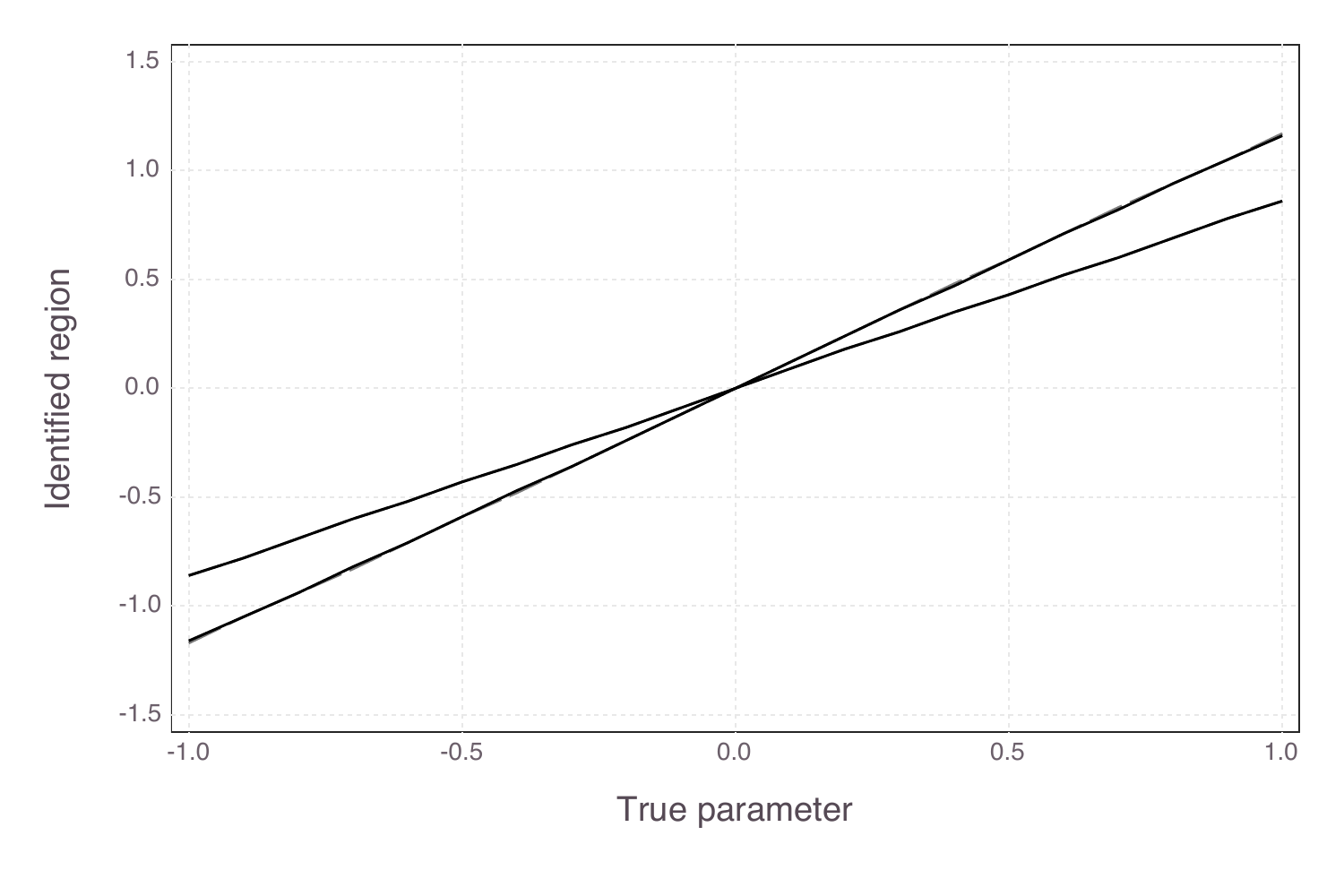} & \includegraphics[width=80mm, height=55mm]{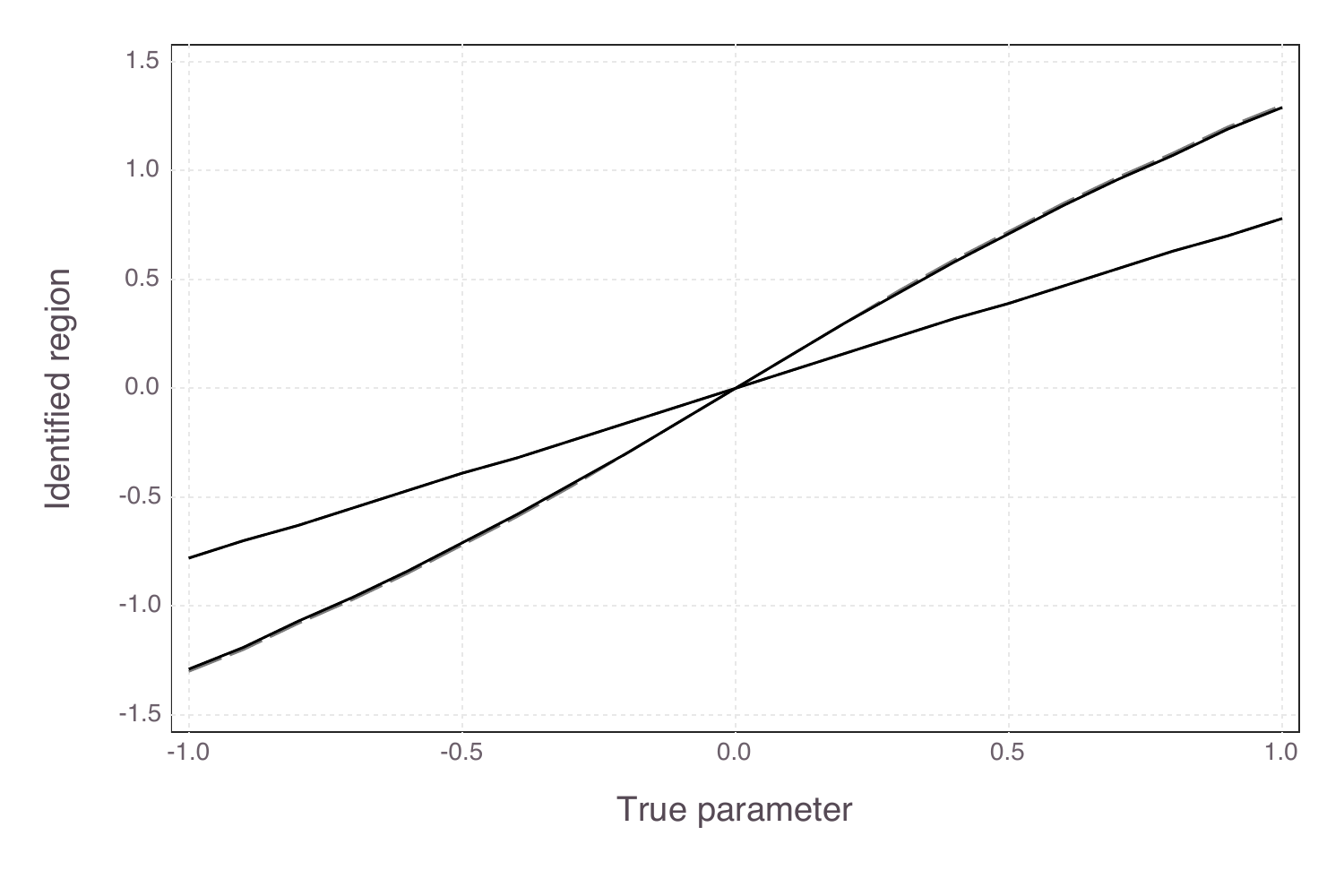}\\				
			\end{tabular}
		\end{center}
		\par
		\textit{{\footnotesize Notes: Approximate upper and lower bounds of the identified set $\Theta^I$ in a logit model (left column) and a probit model (right column) with $T=2$ based on a discretization of unobserved heterogeneity with $K=5,50,500$ support points respectively. The true identified set is depicted by the solid lines while the approximations are indicated by the dashed lines. The population value of $\theta$ is given on the x-axis.}}
	\end{figure}
	
	\clearpage
	
	\begin{figure}[tbp]
		\caption{Approximate identified sets for average partial effects in logit and probit models with $T=2$\label{App_Fig_IS_APE_rob}}
		\begin{center}
			\begin{tabular}{cc}
				LOGIT MODEL & PROBIT MODEL\\
				\multicolumn{2}{c}{$K=5$}   \\
				\includegraphics[width=80mm, height=55mm]{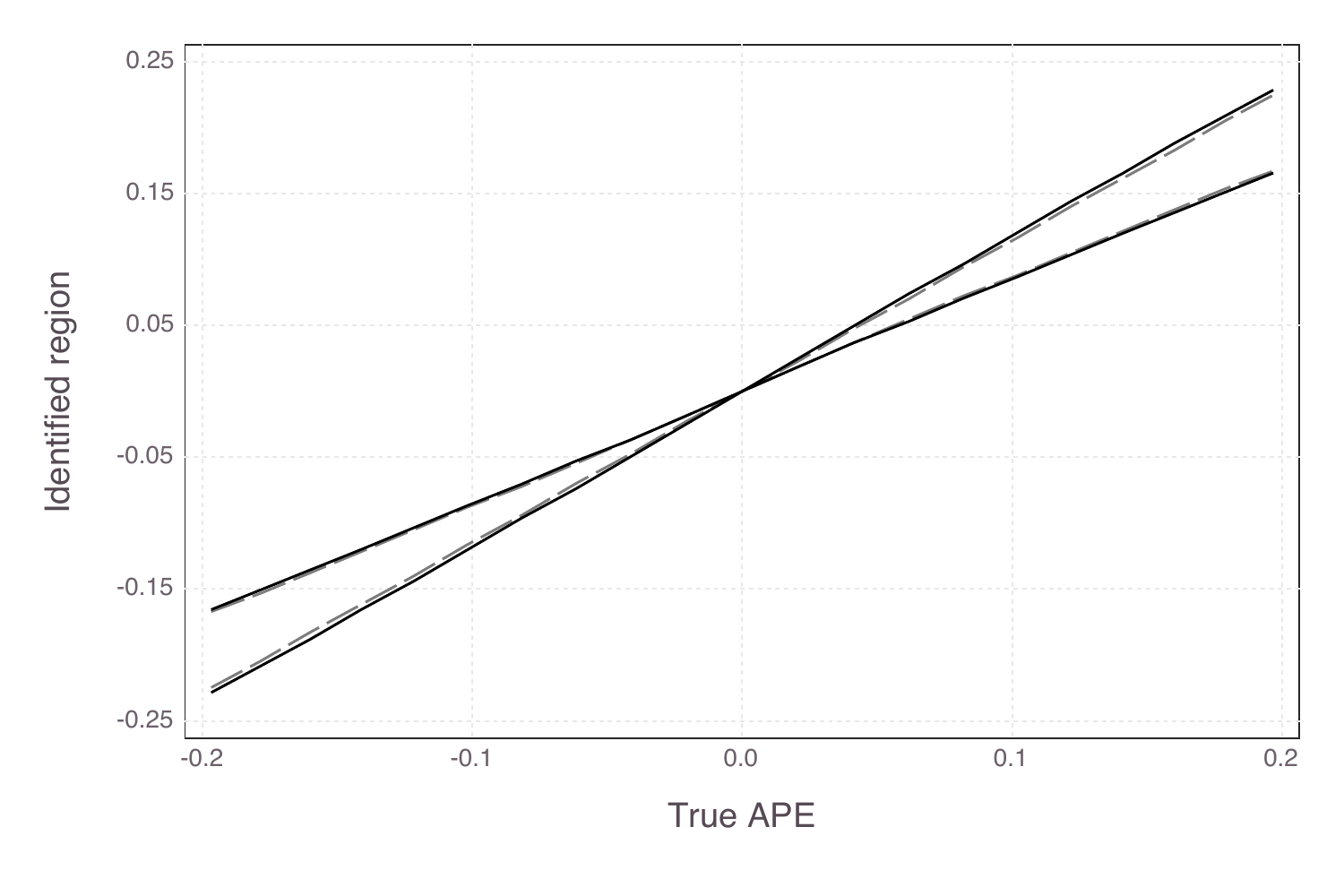} & \includegraphics[width=80mm, height=55mm]{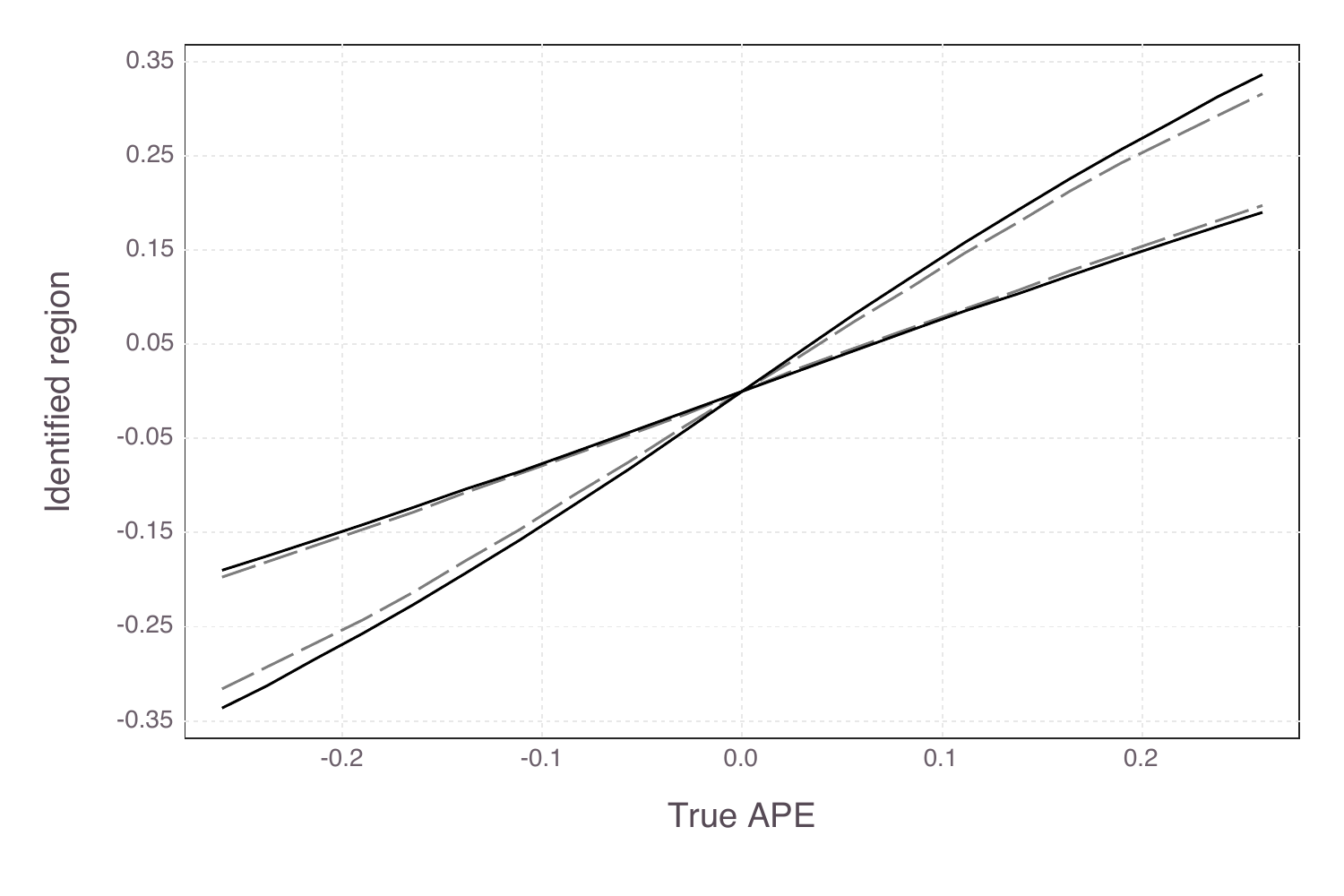} \\
				\multicolumn{2}{c}{$K=50$} \\  
				\includegraphics[width=80mm, height=55mm]{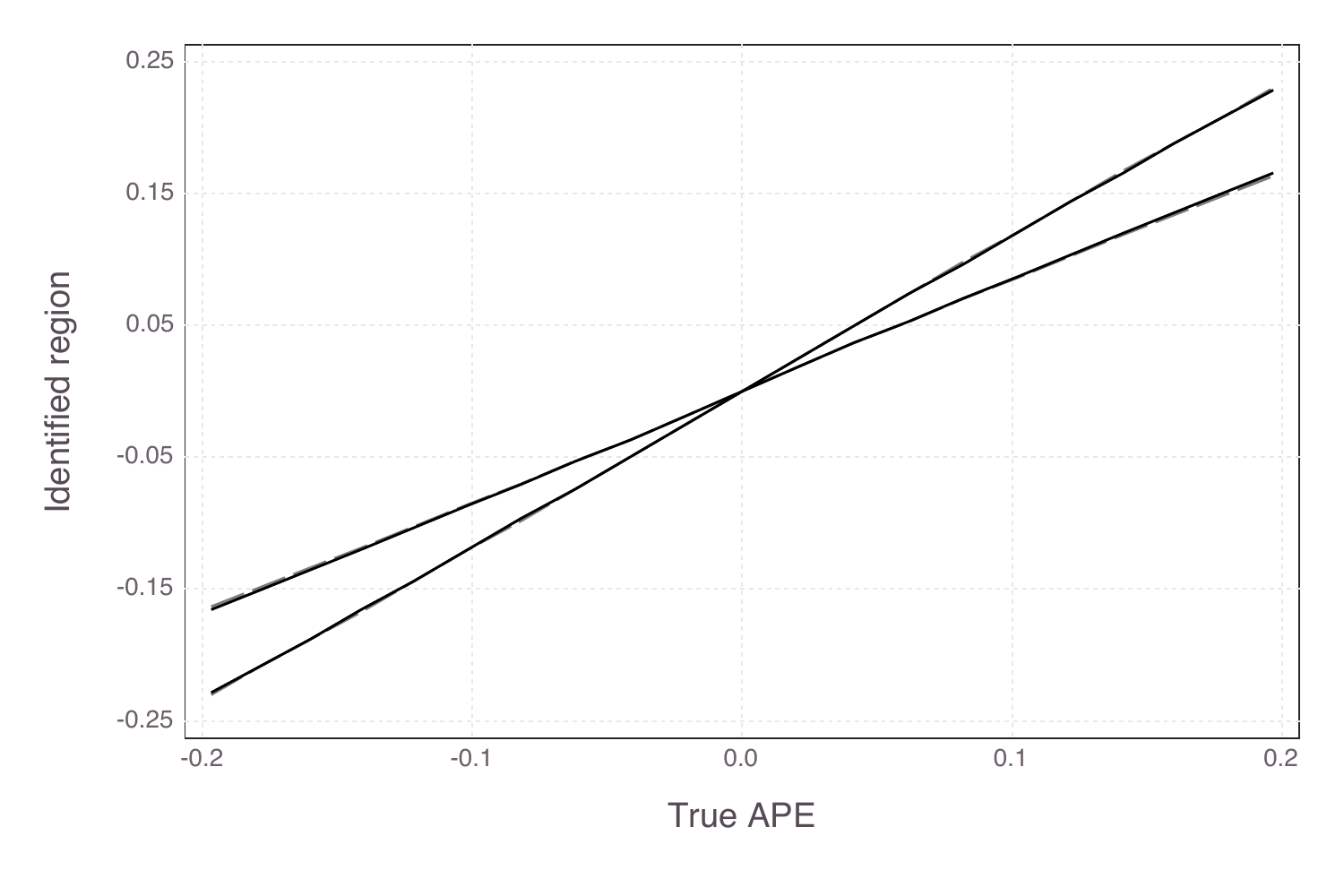} & \includegraphics[width=80mm, height=55mm]{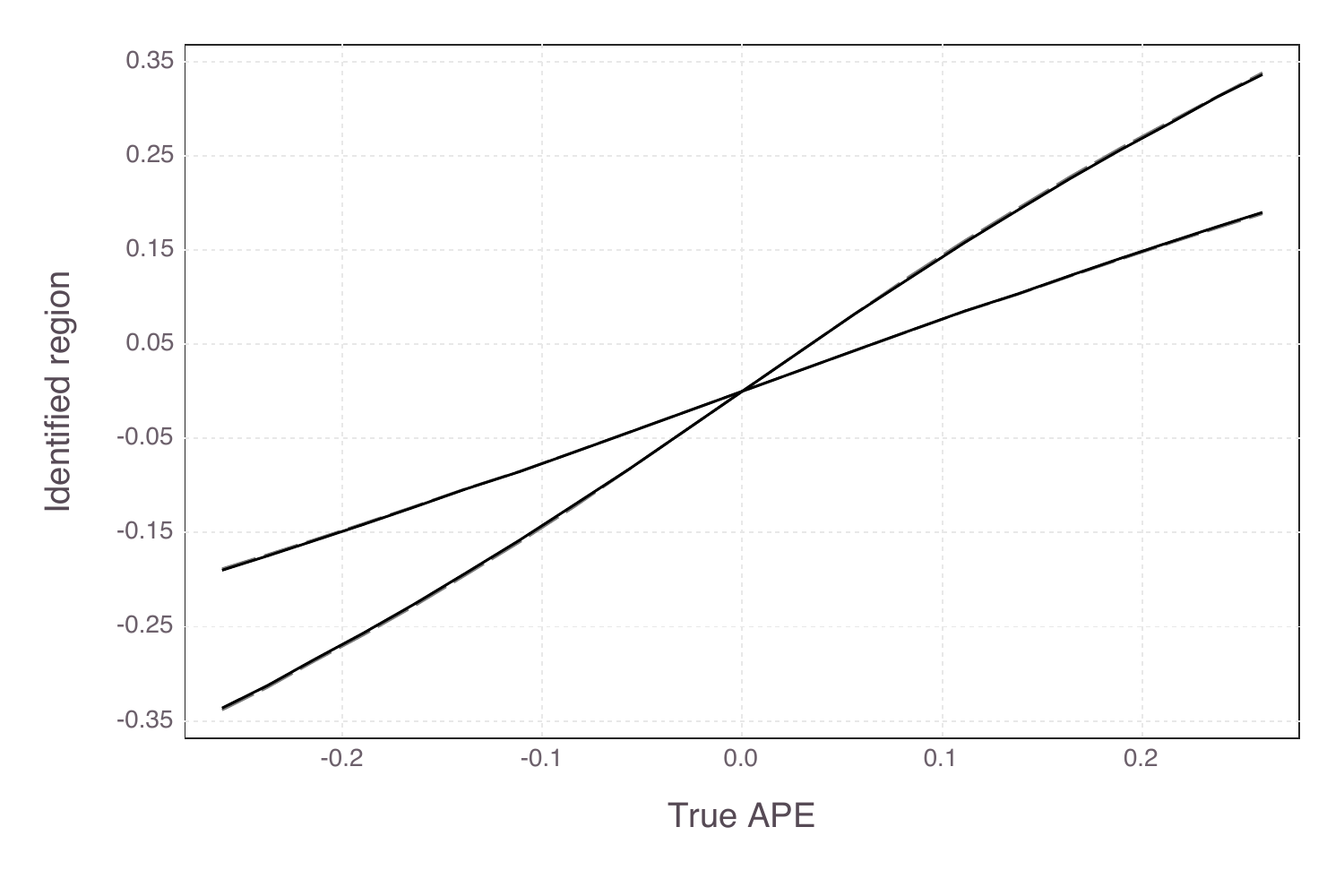} \\
				\multicolumn{2}{c}{$K=500$}\\  
				\includegraphics[width=80mm, height=55mm]{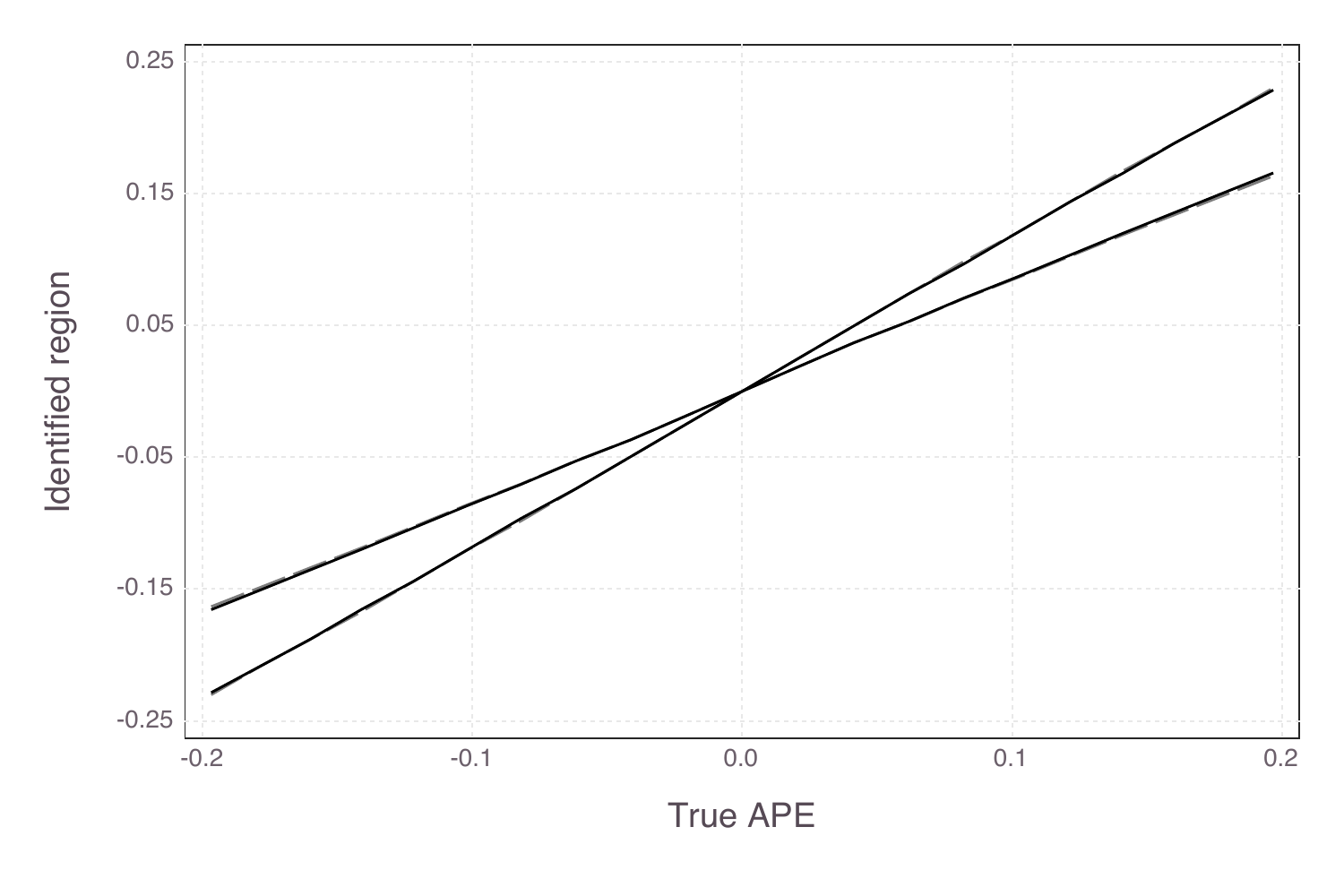} & \includegraphics[width=80mm, height=55mm]{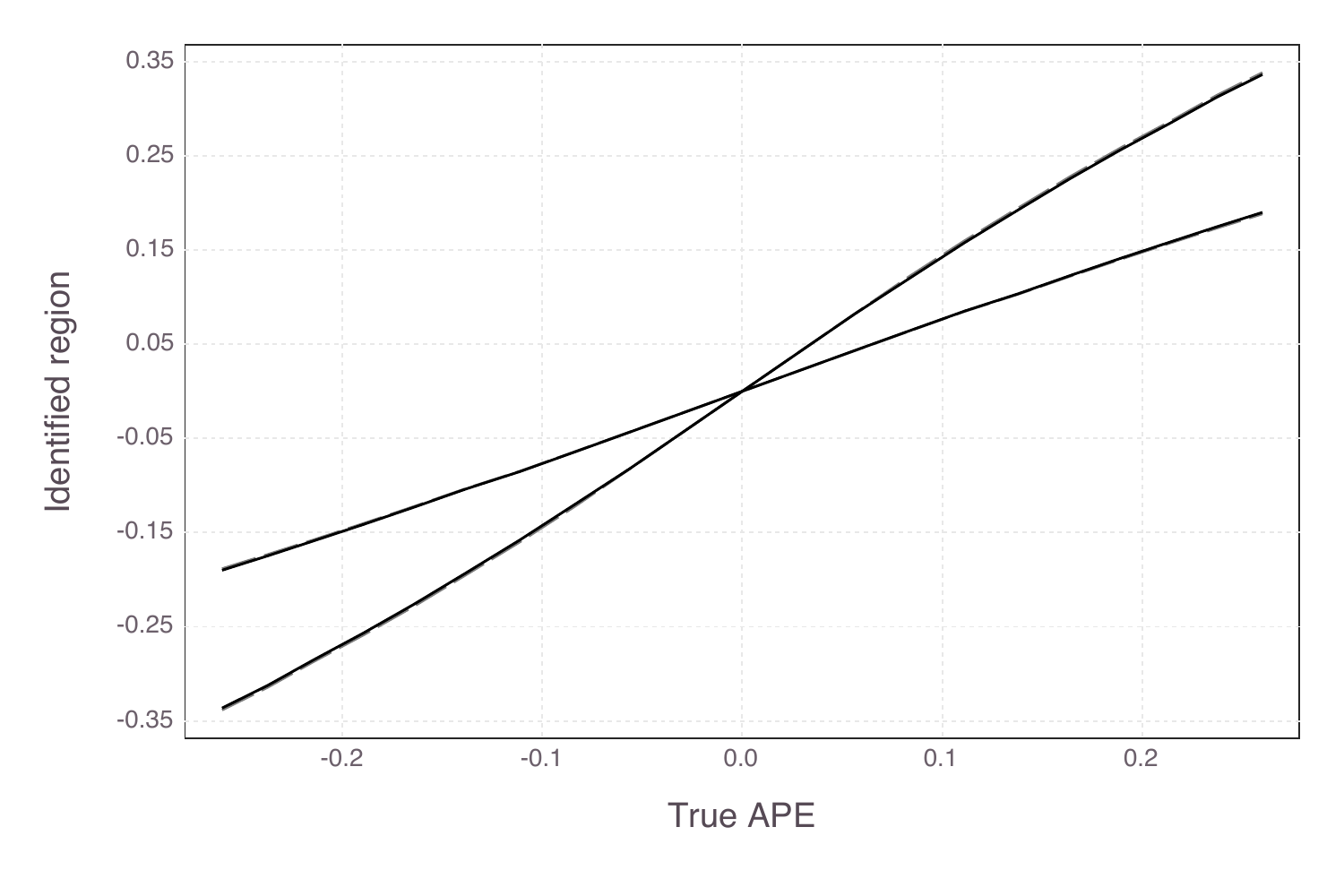}\\				
			\end{tabular}
		\end{center}
		\par
		\textit{{\footnotesize Notes: Approximate upper and lower bounds of the identified set for average partial effects in a logit model (left column) and a probit model (right column) with $T=2$ using a discretization of unobserved heterogeneity with $K=5,50,500$ support points respectively. The true identified set is depicted by the solid lines while the approximations are indicated by the dashed lines. The population value is given on the x-axis.}}
	\end{figure}
	
\end{document}